\pdfoutput=1
\documentclass[iop]{emulateapj}

\usepackage{rotating}
\usepackage{bm}
\usepackage{longtable,tabu}

\shorttitle{Distant RR Lyrae Stars in the Milky Way}
\shortauthors{Medina et al.}

\begin{document}

\title{Discovery of distant RR Lyrae stars in the Milky Way using DECam}

\author{Gustavo E. Medina\altaffilmark{1,2}, Ricardo R. Mu\~noz\altaffilmark{1}, A. Katherina Vivas \altaffilmark{3}, Jeffrey L. Carlin\altaffilmark{4}, Francisco F\"{o}rster\altaffilmark{2,5}, Jorge Mart\'inez\altaffilmark{1,5,2}, Llu\'is Galbany\altaffilmark{6}, Santiago Gonz\'alez-Gait\'an\altaffilmark{5,3,7},  Mario Hamuy\altaffilmark{1,2}, Thomas de Jaeger\altaffilmark{8,2,1}, Juan Carlos Maureira\altaffilmark{5}, Jaime San Mart\'in\altaffilmark{5}
}

\altaffiltext{1}{Departamento de Astronom\'ia, Universidad de Chile, Casilla 36-D, Santiago, Chile.}
\altaffiltext{2}{Millenium Institute of Astrophysics MAS.}
\altaffiltext{3}{Cerro Tololo Inter-American Observatory, National Optical Astronomy Observatory, Casilla 603, La Serena, Chile.}
\altaffiltext{4}{LSST, 950 North Cherry Avenue, Tucson, AZ 85719, USA.}
\altaffiltext{5}{Center for Mathematical Modelling, Universidad de Chile, Av. Blanco Encalada 2120 Piso 7, Santiago, Chile.}
\altaffiltext{6}{PITT PACC, Department of Physics and Astronomy, University of Pittsburgh, Pittsburgh, PA 15260, USA.}
\altaffiltext{7}{CENTRA, Instituto Superio T\'ecnico, Universidade de Lisboa, Portugal}
\altaffiltext{8}{Department of Astronomy, University of California, Berkeley, CA 94720-3411, USA.}

\begin{abstract}

We report the discovery of distant RR Lyrae stars, including the most distant known in the Milky Way, using data taken in the $g-$band with the Dark Energy Camera as part of the High cadence Transient Survey (HiTS; 2014 campaign). 
We detect a total of $173$ RR Lyrae stars over a $\sim 120$\,deg$^2$ area, including both known RR Lyrae and new detections. 
The heliocentric distances $d_{\rm H}$ of the full sample range from $9$ to $>200$\,kpc, with $18$ of them beyond $90$\,kpc.
We identify three sub-groups of RR Lyrae as members of known systems: the Sextans dwarf spheroidal galaxy, for which we report $46$ new discoveries, and the ultra-faint dwarf galaxies Leo~IV and Leo~V.
Following an MCMC methodology, we fit spherical and ellipsoidal profiles of the form $\rho(R)\sim R^n$ to the radial density distribution of RR Lyrae in the Galactic halo.
The best fit corresponds to the spherical case, for which we obtain a simple power-law index of $n=-4.17^{+0.18}_{-0.20}$, consistent with recent studies made with samples covering shorter distances.
The pulsational properties of the outermost RR Lyrae in the sample ($d_{\rm H}>90$\,kpc) differ from the ones in the halo population at closer distances. 
The distribution of the stars in a Period-Amplitude diagram suggest they belong to Oosterhoff-intermediate or Oosterhoff II groups, similar to what is found in the ultra-faint dwarf satellites around the Milky Way. 
The new distant stars discovered represent an important addition to the few existing tracers of the Milky Way potential in the outer halo.

\end{abstract}


\maketitle

\section{Introduction}
The outermost regions of the Milky Way (MW) halo are key probes of the recent assembly history of our Galaxy. Models suggest that stars in the outer halo (beyond Galactocentric radii of $R_{\rm GC} \gtrsim 100$~kpc) likely originated in relatively recently-accreted satellite galaxies \citep[e.g.,][]{BJ05,Zolo09}. 
While current models of galaxy formation generate specific predictions of the amount of stellar substructure in the outermost part of the halo, these have been hard to explore because of the lack of deep, large area surveys of tracers with reliable distance estimates. Because RR Lyrae pulsational variable stars (RRLs) are easily identified in time-series data, are intrinsically bright, and follow well-known period-luminosity relations, they provide a means of mapping the distant halo at distances beyond $d\gtrsim100$~kpc. \citet{Sanderson2017b} predicted that different accretion histories of $12$ synthetic halos \citep[from][]{BJ05} yield populations of $2,000-10,000$ RRLs between $100 < d < 300$~kpc in the MW halo, with roughly half of these in intact dwarf galaxies, and half unbound from their parent satellite.

The accreted dwarfs that are expected to have contributed their stellar populations to the MW halo imprint evidence of the MW's accretion history in the radial stellar density profile. 
Studies of the halo density profile with various tracers \citep[e.g.][summarized later in Table~\ref{tab:powerLaw}]{Saha85, Vivas2006,Wat09,Deas11,Ses11,Akh12} have found widely varying stellar density slopes at large radii, making it difficult to place the MW in a broader context. The discrepancies may be due in part to small samples, especially at large distances from the Galactic center. The addition of outer halo stars is vital to anchor density profile fits in regions of the Galaxy where recent accretion should dominate, and where long dynamical times preserve a record of the accretion events. 
Simulations also suggest that there is a clear difference in the behavior of the inner and outer halo number density profiles, probably driven by different formation processes \citep[see e.g.,][]{BJ05}. The inner halo is thought to contain both accreted and formed in-situ populations, while the stellar component of the outer halo seems to be formed mainly by accretion events \citep{Zinn93,BJ05,Carollo07,Abadi06,Zolo09}.
There is possible evidence for the different formation pathways of the inner/outer halo in the form of a break in the radial number density profile of MW stars near $25$\,kpc from the Galactic center \citep[see e.g.,][]{Saha85, Wat09, Deas11, Ses11}. However, the behavior of the MW's radial density profile at Galactocentric distances ($R_{\rm GC}$) beyond $80$\,kpc is a subject that has not been covered with similar depth, mainly due to incompleteness in current surveys at large distances.
 
Furthermore, the most distant stars are vital tracers for the estimation of the total mass of the MW. Detailed predictions (e.g., the number and luminosity function of satellites) for MW-like galaxies extracted from cosmological models for comparison with observations are highly sensitive to the total mass of the host halo \citep[e.g.,][]{Geha2017}; thus, determining a reliable total mass for our Galaxy becomes essential if one wishes to use it as a cosmological laboratory. Unfortunately, the total mass of the MW within $150$\,kpc is known only within a factor of two \citep[e.g.,][]{Eadie2016, Ablimit2017}.
Mass modeling of the MW halo is most strongly constrained by tracers in the outermost regions; RRLs are thus valuable probes of the Galactic potential because they can be found at large distances, and their distances can be determined to better than $\sim10\%$ with ground-based data.

Regarding remote Milky Way stars, only a small number have been detected at heliocentric distances ($d_{\rm H}$) larger than $100$\,kpc. A summary can be found in \citet{Boch14}. In that study the authors reported the discovery of the two most distant MW stars known to date, with estimated distances larger than $200$\,kpc and classified as M giants. These stars are intrinsically bright which makes them good tracers of halo structure at large distances, but the distance estimations suffer from significant uncertainties (\citealt{Boch14} adopted a $25\%$ of uncertainties). 

Other distance tracers such as pulsational variables have been commonly used to map the Galactic halo over a wide range of distances. RR Lyrae stars, for example, have been identified from a few parsecs to beyond $100$\,kpc \citep[e.g.,][\textrm{reaching} $d_{\rm H}\sim 115$\,kpc]{Wat09, Dra13_100kpc, Sesar2017b}.
RRLs are old, metal-poor pulsating variables that are considered standard candles in the same way as Cepheids, though they are not as luminous.
The light curves of these RRL variables have very characteristic shapes and they are usually classified into two main groups: $ab$ and $c$-type RRLs. The first class (RRab) are fundamental mode pulsators, with saw-tooth shaped light curves and a negative correlation between their amplitudes and periods, which are known to be of the order of $0.6$\,days. RRc, on the other hand, pulsate on the first overtone and display more sinusoidal light curves. The latter have in general shorter periods ($\sim 0.35 $\,days) and smaller amplitudes compared to RRab.

The discovery of groups of RRLs at medium to large distances has become particularly important for the physical description of substructures in the inner and outer halo. Some examples are the discovery and characterization of the Sagittarius stellar tidal stream \citep{Vivas01,Vivas2006,Prior09,Wat09,Ses10,Sesar2017b,Dra13,Zinn14}, the Virgo stellar stream \citep{Duffau06a,Viv16,Sesar2017b}, the Pisces overdensity \citep{Ses10}, 
or the Gemini stream that extends beyond $100$\,kpc \citep{Dra13_100kpc}, among others.
However, the discovery of very distant, isolated stars in the halo brings out questions related to the understanding of their origin. 
Since they are not expected to have formed in the outskirts of the halo, the origin of distant stars is generally thought to be either the gravitational interaction between the Milky Way and its satellites or the ejection from the center (or disk) of the Galaxy \citep{BJ05, Brown05, Zolo09}. Simulations suggest that it is possible to reproduce important stellar halo properties (the break in the density profiles for instance) taking only accretion events into consideration \citep{Deas13}. In this context, an interesting suggestion regarding the origin of distant RRL comes from evidence that all known Milky Way's dwarf galaxies have at least one RRL \citep{Boettcher13, Viv16}. This would make RRL potential tracers of faint satellite systems in the outer halo.
Indeed, based on the data presented in this work, \citet{Medina2017} recognized the presence of the ultra faint dwarf galaxies Leo IV and Leo V based on compact groups of distant RRLs. 

To date, large RRLs catalogs have been constructed using data from different variability surveys that map different parts of the halo. 
Among them, the Quasar Equatorial Survey Team (QUEST) RRL catalog \citep{Viv04} contains $457$ objects with $V<19.5$, and subsequently with La Silla QUEST Southern Hemisphere Variability Survey (LSQ) came the discovery of $1013$ RRab and $359$ RRc distributed across $\sim 840 \ \rm{deg}^2$ of the sky in the range $150^{\circ} < \rm{RA} < 210^{\circ}$ and $-10^{\circ} < \rm{Dec} < 10^{\circ}$ \citep{Zinn14}. The latter contains stars with $d_{\rm H}$ between $5$ and $80$ kpc. 
More recently, the Catalina Surveys identified $\sim 43,500$ RRLs over $\sim 30,000$\,sq degrees of the sky up to heliocentric distances of $\sim 60-110$\,kpc \citep{Drake17}.
The Pan-STARRS1 (PS1) $3\pi$ survey has proven to be also an excellent source for RRLs, despite its poor temporal sampling, by identifying a sample of $\sim 45,000$ RRLs up to $\sim 130$\,kpc from the Sun with a $90\%$ purity \citep{Sesar2017a}. All these surveys have found several distant RRLs, but the surveys' low completenesses at faint magnitudes limit the depth of the findings.

In this contribution, we use data from the High Cadence Transient Survey \citep[HiTS;][]{Forster16} to search for distant RRL (beyond $100$\,kpc). HiTS is a survey designed for the detection of young supernovae events with a total sky coverage of $\sim 350$ square degrees and a limiting $g$ magnitude that varies between 23 and 24.5 \citep{Forster16}. Since the cadence and observing strategy of the survey are well matched for RRL detection, it should allow us to find distant RRL in the observed region, if present.
Here we discuss results from $\sim120$~deg$^2$ of HiTS survey data, corresponding to the second campaign which was held during semester 2014A.

\begin{figure*}[!t]
\includegraphics[angle=0,width=1.1\textwidth]{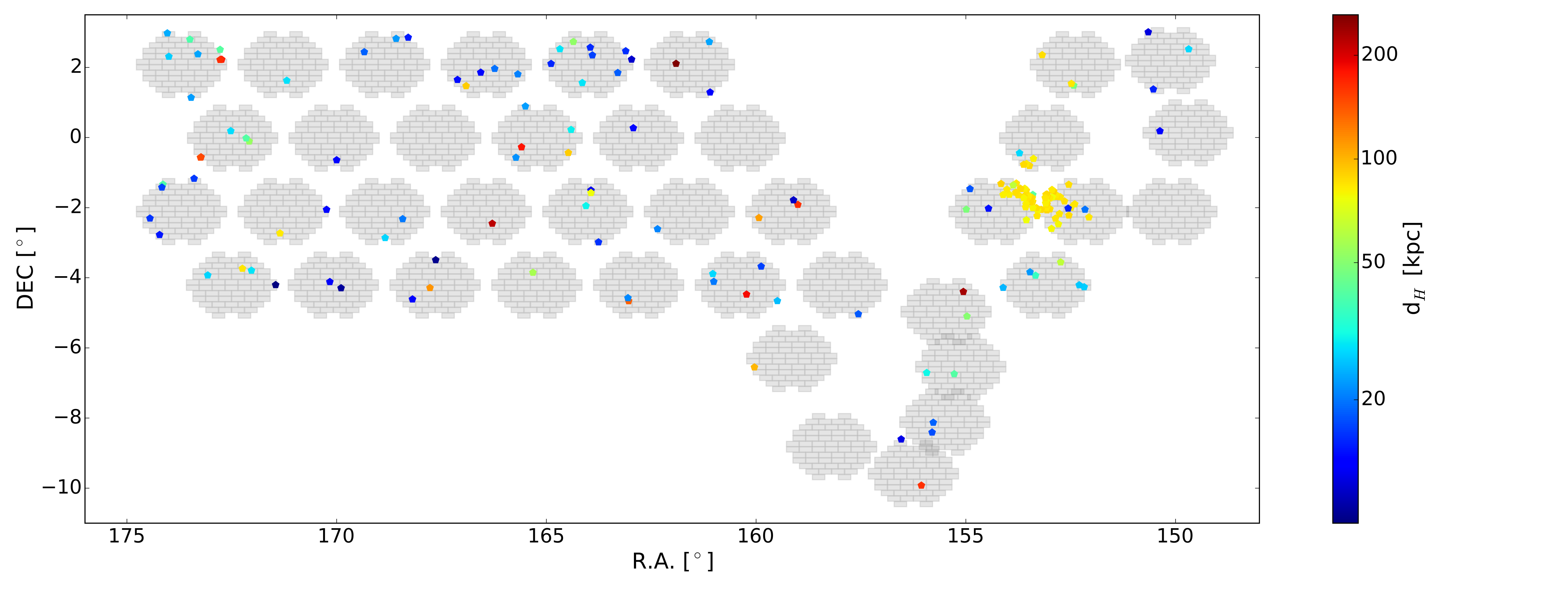}
\caption{Color coded plot of the distribution of our $173$ RRL stars in the sky, with the colors representing their heliocentric distances. An approximation of the HiTS footprint is shown in grey in the background as a reference.}
\label{fig:colordistancias}
\end{figure*}

The structure of this paper is organized as follows: In section 2 an overview of the observation context of this work is given, including details of HiTS. In \S3 the candidates selection and characterization process is described, as well as the distance determination. The analysis of the sample, including the presence of substructures in our data like the Sextans dwarf spheroidal galaxy and the description of the most distant candidates is presented in \S4 and \S5, respectively. In \S6 we use our sample to construct radial density profiles of the halo and compare them with previous studies. Finally, in \S7 our main results are summarized.

\section{Observations}

HiTS is a deep optical campaign carried out with the Dark Energy Camera \citep[DECam,][]{Flaugher15} mounted at the prime focus of the Blanco $4$\,m telescope at Cerro Tololo Inter-American Observatory (CTIO). The camera contains $62$ CCDs with $520$ megapixels and generates three square degree images. It is currently the largest etendue camera in operation in the southern hemisphere, and it will remain so until the Large Synoptic Survey Telescope (LSST) begins operations in $2022$.

One of the main goals of HiTS is the detection of optical transient objects, in particular young supernovae events mainly in the $g$ SDSS photometric system filter ($\sim 4000$ to $5500$~\AA) in the search for empirical evidence of the shock breakout phenomena. The fact that the survey can detect transient objects with characteristic timescales of hours make HiTS suitable for RRLs studies as well. 

The observations used for the current work took place between UT 2014 February 28 and March 4 (five consecutive nights). Forty fields were observed with a cadence of two hours, four times per night in the $g-$band and the exposure times varied from 160 seconds ($83\%$ of the observations) to 173 and 174 seconds ($3\%$ and $14\%$ respectively). This translates into 20 epochs per field, with the exception of one field (centered in $151.5^{\circ}$ RA, $-2^{\circ}$  DEC), which has 37 (see \citealt{Forster16} for detailed field positions). The sky was clear during all nights, with typical seeing values around $1.5$\,arcseconds. 

The fields observed by HiTS were selected almost entirely blindly, covering $\sim 120$ square degrees in the sky region from $150^{\circ} < \rm{RA} < 175^{\circ}$ and $-10^{\circ} < \rm{Dec} < 3^{\circ}$ (see Figure~\ref{fig:colordistancias}).  The only requirement for the fields was the observing availability during the entire night. Six of the fields were chosen for being known cluster fields and two for being SuperCOSMOS fields \citep{Hambly2001}. The data obtained were stored on the National Laboratory for High Performance Computing (NLHPC) cluster at the Center for Mathematical Modeling (CMM) of Universidad de Chile and then analyzed using the HiTS pipeline. 

The HiTS pipeline is based on basic calibrations, image substraction, candidate filtering and visualization \citep{Forster16}. The instrumental signature removal steps were performed using the DECam community pipeline \citep{Valdes2014}.
For this work, we generated catalogs using the SExtractor photometry software \citep{Bert96}, with a limiting apparent magnitude of 23--24.5 \citep{Forster16}.

\section{SEARCH AND CHARACTERIZATION OF RRLs}
\subsection{Photometric Calibration}

\begin{figure}
\includegraphics[angle=0,scale=.40]{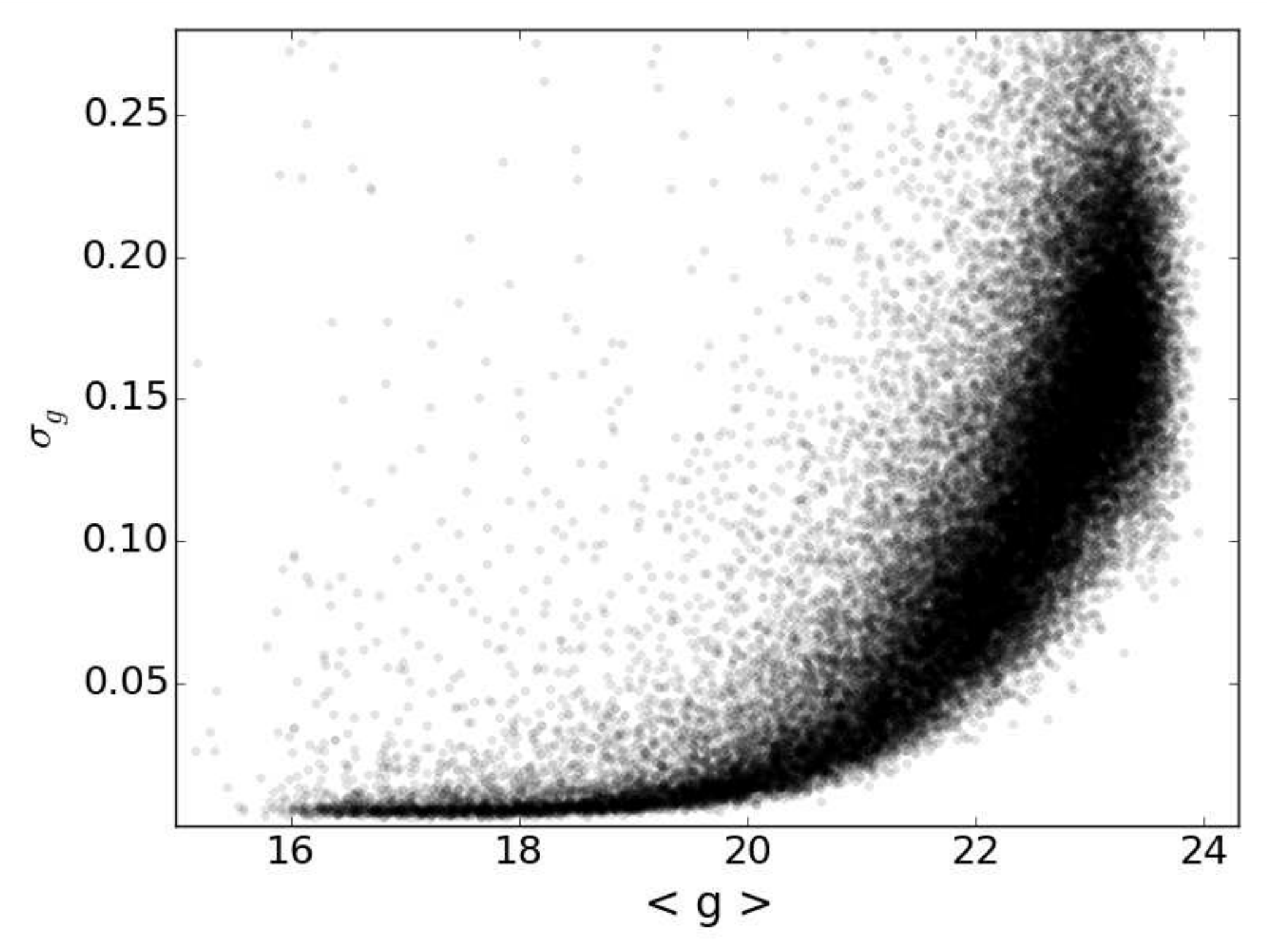}
\caption{Variation of $\sigma_g$ with the average magnitude for one of the HiTS fields (black dots). To model the data, a cubic spline interpolation was applied to the sigma-clipped data.}
\label{fig:stdmags}
\end{figure}

In order to select periodic variables we need to
define a common $x, y$ pixel coordinate system using the output catalogs generated by SExtractor,
a process we call alignment. 
We select the second epoch observation as the reference frame for having the best observing conditions.
The scaling constants needed to do the alignment were found by the HiTS pipeline \citep{Forster16}. Subsequently, we performed a cross-match between the aligned catalogs and rejected as possible candidates sources with fewer than five detections. 
Sources whose mean flux uncertainties are larger than two times the flux standard deviation were also filtered out. 
Pixel coordinates were transformed into equatorial coordinates following the procedure described in \citet{Forster16}. 

To account for the effects of the atmospheric conditions over the different epochs, we calculated a relative zero-point associated with the reference. This was performed comparing the instrumental magnitudes given by

\begin{equation}
g_{\rm inst} = -2.5\, \log \left(\frac{\rm counts}{s}\right) - a_g - k_g\, A
\label{eq:ginst}
\end{equation}

\noindent where A is the airmass, $a_g$ is the photometric zero-point in the $g-$band (one for each CCD) and $k_g$ is the first-order extinction in $g$\footnote{\url{http://www.ctio.noao.edu/noao/content/Mean-Photometric-Standard-Star-Module-PSM-Solutions-mean-zeropoints-color-terms-extinctions}}. 
Then, the $g$ magnitude was calculated as following:

\begin{equation}
\begin{array}{lc}
g_{\rm ref} = g_{\rm inst} + \Delta_{\rm rel}\\
g = g_{\rm ref} + \Delta_{\rm PS}
\end{array}
\label{eq:gs}
\end{equation}

In the previous equation, $\Delta_{\rm rel}$ refers to the relative zero-point between the different epochs and the reference, which was calculated on a chip-by-chip basis, and $\Delta_{\rm PS}$ is a calibration zero-point with respect to the PS1 DR1 public catalog. To obtain the PS1 zero-points, we compared the instrumental magnitudes for all the stars in a given chip (corrected by $\Delta_{\rm rel}$) with the corresponding magnitudes listed in the PS1 database. This yields $60$ different zero-points for a given field. We repeated this procedure for all fields in the PS1 footprint.

To estimate the photometric uncertainties, we assume that the uncertainties for each star magnitude can be obtained using a non-parametric model of the standard deviation and mean magnitude relation for all the stars in their respective fields, as can be seen in Figure~\ref{fig:stdmags}.
We do not use the flux errors given by SExtractor since they are, in general, underestimated 
\citep[][ApJ, submitted]{Gawiser2006, Martinez2018} and therefore are not reliable when accounting for the photometric uncertainties.
The photometric uncertainties reach $\sim 0.10$ mag at $g\sim 22$ and $\sim 0.15$ mag at $g\sim 23$. 

The final magnitudes include extinction corrections for which the re-calibrated dust maps of \citet{Sch11} were used. The extinction values for the $g-$band were calculated using the relation $A_g = 3.303 \cdot E(B-V)$ from the mentioned work. Following this, the average $A_g$ of the RRLs in the HiTS fields is $\sim 0.14$ mag, with a standard deviation of $0.03$ mag. 

\subsection{Selection of RRLs}

For period determination, we ran the generalized version of the Lomb-Scargle periodogram \citep[GLS;][]{Zech09}, which provides more accurate frequencies and is less susceptible to aliasing than the Standard Lomb-Scargle \citep[LS;][]{Lomb76, Scar82} periodogram analysis. We selected as pre-candidates objects with periods greater than $4.8$\,hours ($0.2$\,days) and less than $21.6$\,hours ($0.9$\,days), and with a GLS statistical level detection of $< 0.08$. The statistics were computed by using the GLS tool from the astroML python module \citep{Vander12}. In spite of the reduction in the aliasing by the GLS near certain problematic periods ($0.33$ and $0.50$\,d) compared with the standard LS, we considered it reasonable to filter out objects with periods between $0.32$ and $0.34$\,days, and within $0.49$ and $0.51$ to reduce the number of spurious variables. This resulted in the rejection of $\sim 3\%$ of non-variable sources that were spuriously identified as variables due to period aliasing. In the case where two or more significant periods met the requirements discussed above, we allowed the two most significant to be selected. 
This results in the selection of $2434$ objects.

Finally, objects with $\Delta$ mag $< 0.2$ were filtered out and we use visual inspection of the phased light curves (looking for periods, amplitudes and light-curve shapes typical of RRLs) to obtain our final list of RRLs comprised of only $173$ objects ($\sim 7 \%$). 
A small number of stars look clearly periodic but they do not have the correct amplitude, period or shape for being considered RRLs and they are not further discussed here.
An additional visual inspection of the reference images for each of the RRLs was performed to reject possible extended objects.

Color information provides additional constraints at the moment of selecting/confirming RRLs \citep[see e.g.][]{Ive05}. Unfortunately, HiTS observations were done only in the $g$ band. Not all our stars have a counterpart in SDSS because some of our fields are outside the footprint of that survey, and our limiting magnitude is significantly fainter than SDSS. However, 139 of our RRLs have a counterpart in SDSS DR12 and the ones with small photometric errors in SDSS are constrained within $0.9 < u - g < 1.8$ and $-0.1 < g - r < 0.3$, as expected for RRLs. We also look at the colors in PS1 which completely covers our region although again, does not go as deep as HiTS. In this case we found a match for 161 RRLs and all of them have colors $g - r$, $r - i$ and $i - z$ consistent with RRLs, and confirming our selection criteria based on the $g$ light curves is good enough to isolate RRLs.

\begin{figure}
\includegraphics[angle=0,scale=.40]{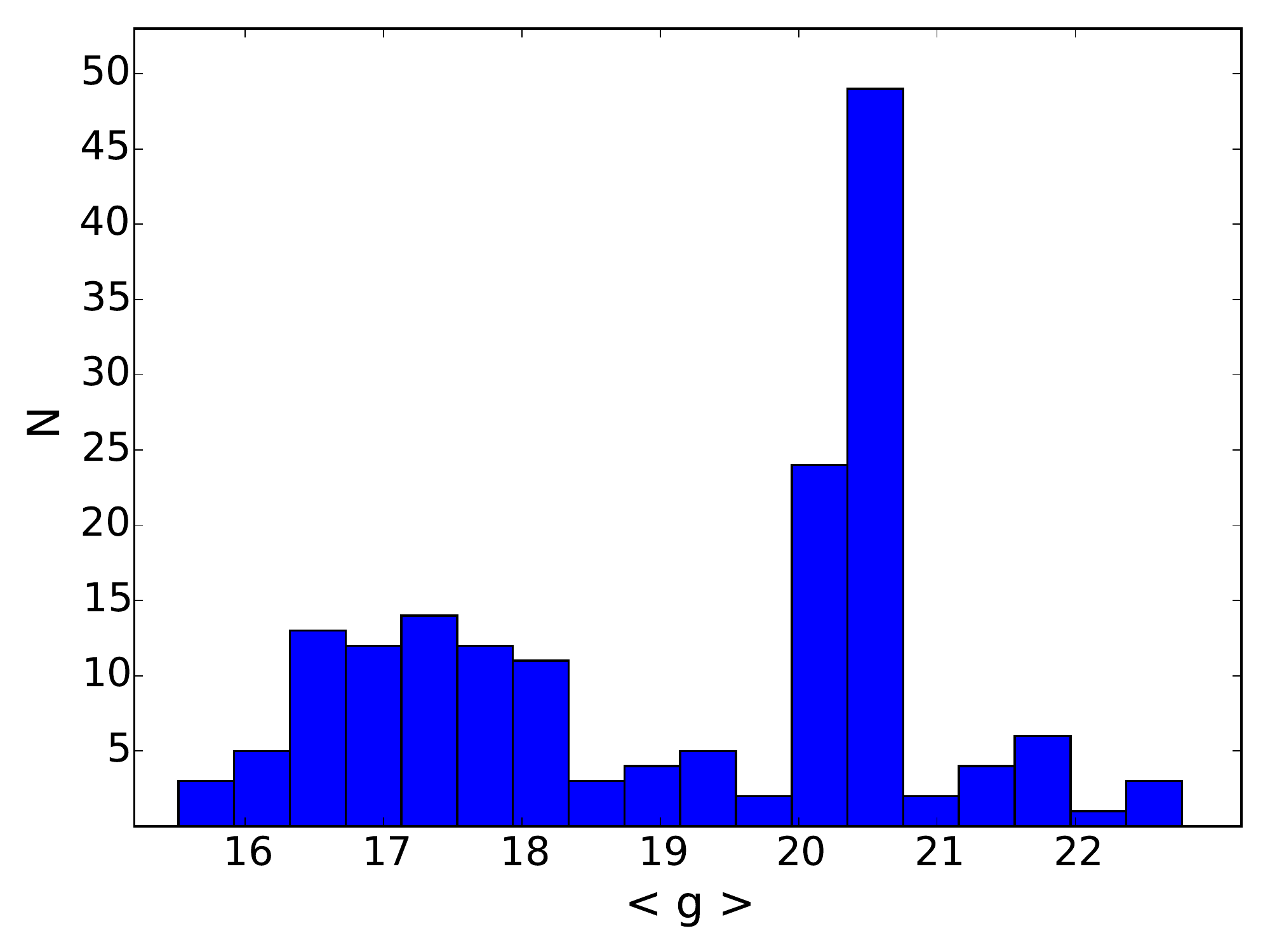}
\caption{Distribution of the mean magnitudes $g$ for the complete sample of RRLs.}
\label{fig:histogram_g}
\end{figure}

To have a more robust method for the estimation of the light curve parameters we adjusted templates of RRLs from  the work by \citet{Ses10}, which was based on SDSS Stripe 82 RRLs. In that study, a list of $23$ templates is provided for the $g-$band, two of which are for RRc. The selection of the best fit for each candidate was based on $\chi^2$ minimization. For the fitting we allowed small variations around the observed amplitude and maximum magnitude ($0.2$ magnitudes for each) as well as for the period and initial phase obtained through GLS ($0.01$\,days and $0.2$, respectively). 

The 173 RRLs ($131$ RRab and $42$ RRc) have mean magnitudes between $15.5$ and $22.7$.
Figure~\ref{fig:histogram_g} shows the distribution of the mean $g$ magnitude for the full sample. The strong peak seen at $g\sim 20.5$ correspond to stars associated with the Sextans dSph galaxy and are further discussed in \S4. Surprisingly, there is a non-negligible number of RRLs with $g>21$ which correspond to very distant halo object. 
Folded light curves and tables with the properties of the RRLs are shown in \S5 (for the very distant stars) and in the Appendix material (for the rest of the stars).

\begin{figure}
\includegraphics[angle=0,scale=.40]{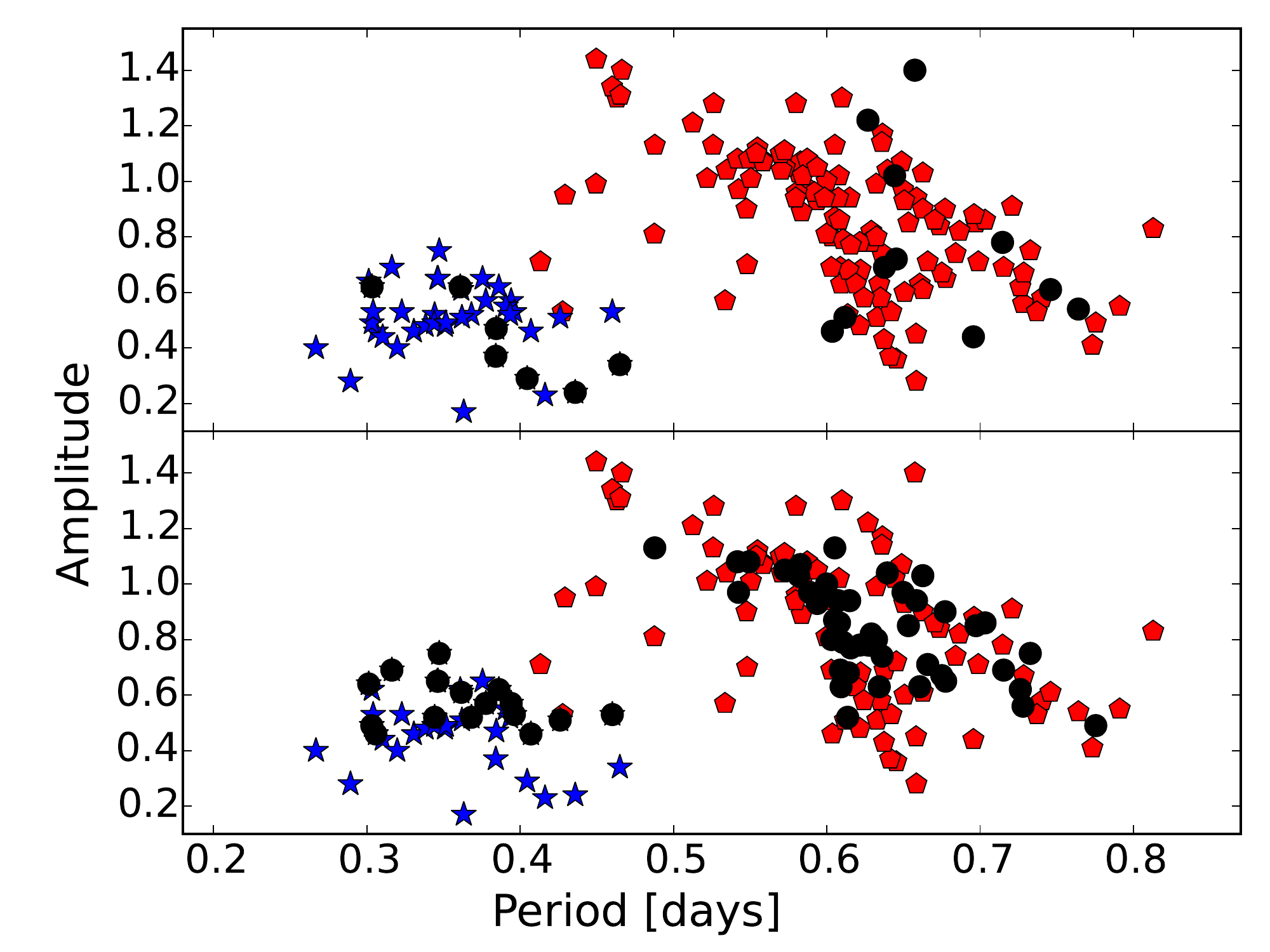}
\caption{Period-Amplitude diagram of the complete sample of RRLs. The blue stars symbols represent RRc stars, while the RRab are plotted with red pentagons. Black filled circles represent RRLs with $d_{\rm H}>90$\,kpc (\it{top}\rm) and RRLs members of the Sextans dSph galaxy (\it{bottom}\rm).}
\label{fig:peramp}
\end{figure}

Figure~\ref{fig:peramp} shows the location of the RRLs in the Period-Amplitude diagram. As expected, the sequences of the Oosterhoff (Oo) groups I and II \citep{Oosterhoff1939,Catelan2015} are clearly visible in this plot, especially for the halo RRab at distances $<90$ kpc (red symbols in Figure~\ref{fig:peramp}). This is reassuring that our methodology is correctly recovering properties of RRLs. We discuss implications of the location in this diagram of the newly found RRLs later in the article.

\subsection{Detection Efficiency}

\begin{figure}
\includegraphics[angle=0,scale=.40]{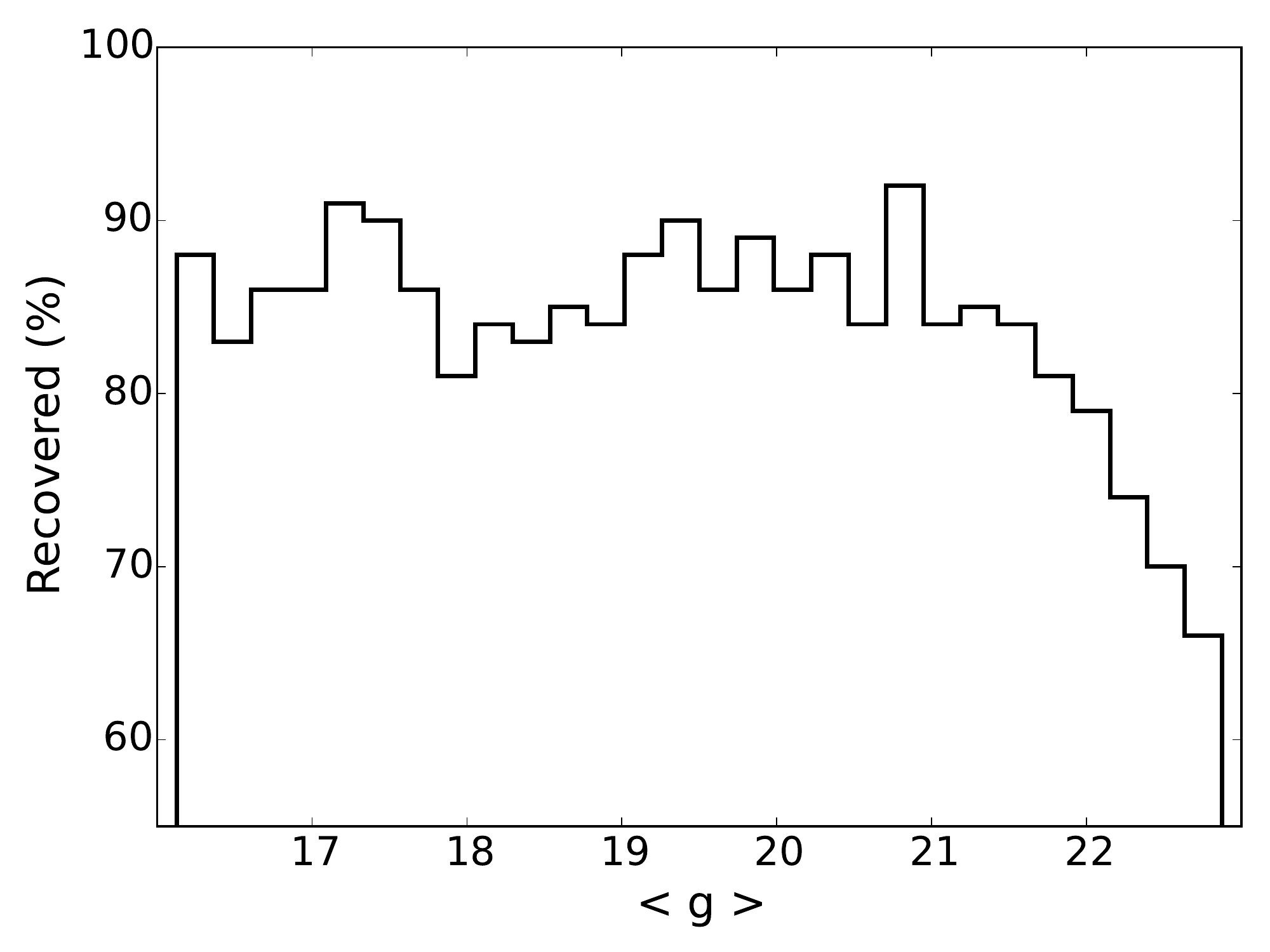}
\caption{Recovery rates as a function of mean $g$ magnitude of synthetic RRLs.}
\label{fig:completitud}
\end{figure}

To estimate the detection efficiency of our survey we generated $5000$ artificial light curves of RRLs mimicking our real cadence and photometric errors. Specifically, the number of observations per object were set equal to the ones of randomly selected stars in our survey. Thus, the number of epochs in the artificial light curves vary from $16$ to $37$. Each light curve was modeled using the parameters (amplitude in $g$, period, template) from the list of SDSS RRLs in \citet{Ses10}, which includes stars with periods within the ranges rejected by our selection criteria.
A random initial phase was added to each simulated star. Finally, photometric errors were added to the template magnitudes according to the values of the main locus of stars seen in Figure~\ref{fig:stdmags}.
The artificial light curves were then processed by our software and stars which pass all our criteria and have a period within 10\% of the real one were flagged as recovered. 

Figure~\ref{fig:completitud} shows the results of the process represented as a histogram where $0.25$ magnitude bins were used. In the bright end, our method is able to recover $\sim 86\%$ of the full sample, and the rate ranges between $81\%$ and $92\%$ down to $g=21.5$. The detection efficiency drops to less than $70\%$ for RRL fainter than $g=22$. 
The estimation of the RRLs detection efficiency does not take the photometric completeness of the survey into account.
A detailed analysis of the photometric completeness of HiTS can be found in \citet{Forster16}.

\subsection{Comparison with previous surveys}

Another way to investigate how many RRLs we are missing in our survey is to compare our results with recent large-area surveys that covered the same portion of the sky as HiTS. In this work we compare with data from the Catalina Real-Time Transient Survey (CRTS) data release 1 (\citealt{Dra13, Dra14}), from the La Silla-QUEST (LSQ) survey \citep{Zinn14} (both have a large number of observations in the $V-$band) and from PS1 ($grizy$ filters but only a few epochs in each band). The overlap in sky coverage is complete in the case of the CRTS and PS1, and partial for LSQ.

When compared with \citet{Dra14}, 
$93$ RRL ($76$ RRab and $17$ RRc) fall into the fields observed by HiTS with magnitudes fainter than $16$ (near our saturation limit)
that we should have been able to detect.
Of these $93$, $69$ matched up with our list within $2$\,arcseconds ($74\%$), and $74$ within $2.5$\,arcseconds ($80\%$). 
In terms of percentages, $74\%$ ($79\%$ for $2.5$\,arcsecs) of the ab types and $76\%$ ($82\%$) of the RRc's were recovered. There are no differences in the classifications they give to the RRLs and what we find in this work. The periods obtained for our sample are in good agreement with the periods for the stars in common, with a mean discrepancy of $2\times 10^{-3} \pm 2\times 10^{-2}$\,days.
We considered the position of the $24$ ($19$) missing RRLs 
in the CCDs and found that at least $7$ of them fell near the edges for the reference frame and therefore were not cleanly detected by our procedure. 
Removing these $7$ stars from the list of potential matches with \citet{Dra14} increases the rate of recovery to $80\%$ ($84\%$) which is closer to the numbers obtained for the theoretical recovery rate computation.
In general, the missed RRLs are relatively bright sources ($V<19$, $d_{\rm H}<40$\,kpc), with $\left<V\right>\sim 17$ and enough phase coverage to perform a good characterization in the CRTS ($N>180$). 
In our sample we classified as RRLs $103$ sources that were not present neither in \citet{Dra13} nor \citet{Dra14}, $74$ of them are RRab, and the remaining $29$ are RRc. These stars may have been missed by these surveys for different reasons, but the most likely is attributable to the different depths of our surveys.
Both \citet{Dra13} and \citet{Dra14} have shallower limiting magnitudes, and have high completeness levels until $V_{\rm CSS}\sim 19.5$.
More than a half of the new RRLs ($\sim 60\%$) are grouped together and seem to be part of the Sextans dwarf spheroidal galaxy (as described in section \S5), while from the remaining objects, $22$ have magnitudes $g > 20$, mostly beyond the detection limits of the CRTS.

In the case of LSQ, which has $43$ RRab and $7$ RRc fainter than $g = 16$, the comparison yielded $44$ stars recovered ($88\%$; 5 RRc and $39$ RRab).
In general, the periods seem to coincide with an absolute mean difference of $4.7\times 10^{-2}$\,days. However, there is one star with a significantly different period (HiTS101413-004502). If we do not include HiTS101413-004502 in the comparison, the period discrepancy turns out to be $|\Delta P|\sim5\times 10^{-4}$.
For that star we found a period of $0.386$\,days as the most likely period, while LSQ gives $0.628$\,days. However, the latter corresponds to the second most probable period according to our procedure. Since LSQ has $87$ observations for that star and we only have $21$, their period is likely more reliable. This discrepant period determination led us to a misclassification of that RRL (from a RRab to a RRc). In Table~\ref{tab:sextans}, the period obtained with our methodology is presented.
Of the RRLs we missed, 4 are RRab and 2 are RRc. Nevertheless, one RRab (LSQ250) fell relatively close to the edges of its CCD, following the criteria applied in the comparison with the CRTS. 
Rejecting LSQ250, the recovery rate rises up to $\sim 90\%$.
As in the case of the comparison with CRTS, we report numerous RRLs that do not appear in the LSQ catalog. We found $93$ new variables in the common regions ($70$ ab's and $23$ c's) most of which seem to be related, again, to the Sextans dSph RRLs overdensity ($65\%$).

The PS1 public catalog has $24.000$ RRLs and was built based on machine learning techniques \citep{Sesar2017a}. 
From this sample, $298$ fall into the HiTS fields.
Based on their calculations, an RRab/RRc score in their selection higher than $0.90$ gives a purity of $0.97$/$0.94$ and a completeness of $0.88$/$0.57$ for $r < 18.5$. If only RRLs with scores higher than $0.90$ are considered, the number of PS1 RRLs in the HiTS fields with mean magnitudes in $g$ fainter than $16$ is $110$. 
A comparison of this group with HiTS RRLs yields $87$ matches within $2$\,arcseconds ($75$ ab's, $12$ c's), i.e., we recover $80\%$ of the intersecting PS1 sample. 
The mean period difference is $0.001$\,d, with $0.07$\,d as the most significant discrepancy, and the mean magnitude difference is $0.09$.
The faintest common RRL, HiTS101243+022118, has $g=20.61$ ($87$\,kpc) in the HiTS catalog, and $g=20.79$ in PS1. 
The classifications of all the stars in this subsample of the PS1 catalog are in agreement with the classifications presented in this work.

It is worth noticing that five of the RRLs in PS1 that were not detected by us fall near the edges of the CCDs for the reference frame. If we do not take into account these RRLs, our recovery rate go as high as $83\%$, in agreement with our estimations.
The rest of the undetected RRL correspond to relatively bright sources, with a mean value of $g$ of $18.3$.

Of the $86$ RRLs that are not clearly detected by PS1, $28$ have $g<20$ ($18$ ab's, $10$ c's), which corresponds to $d_{\rm H} \lesssim 70$\,kpc.
The rest of the stars not present in the PS1 catalog are mostly $ab-$type ($65\%$), and include the entire sample of faint RRLs described in Section 5.

\subsection{Heliocentric distance determination}

\begin{figure}
\includegraphics[angle=19,scale=.50]{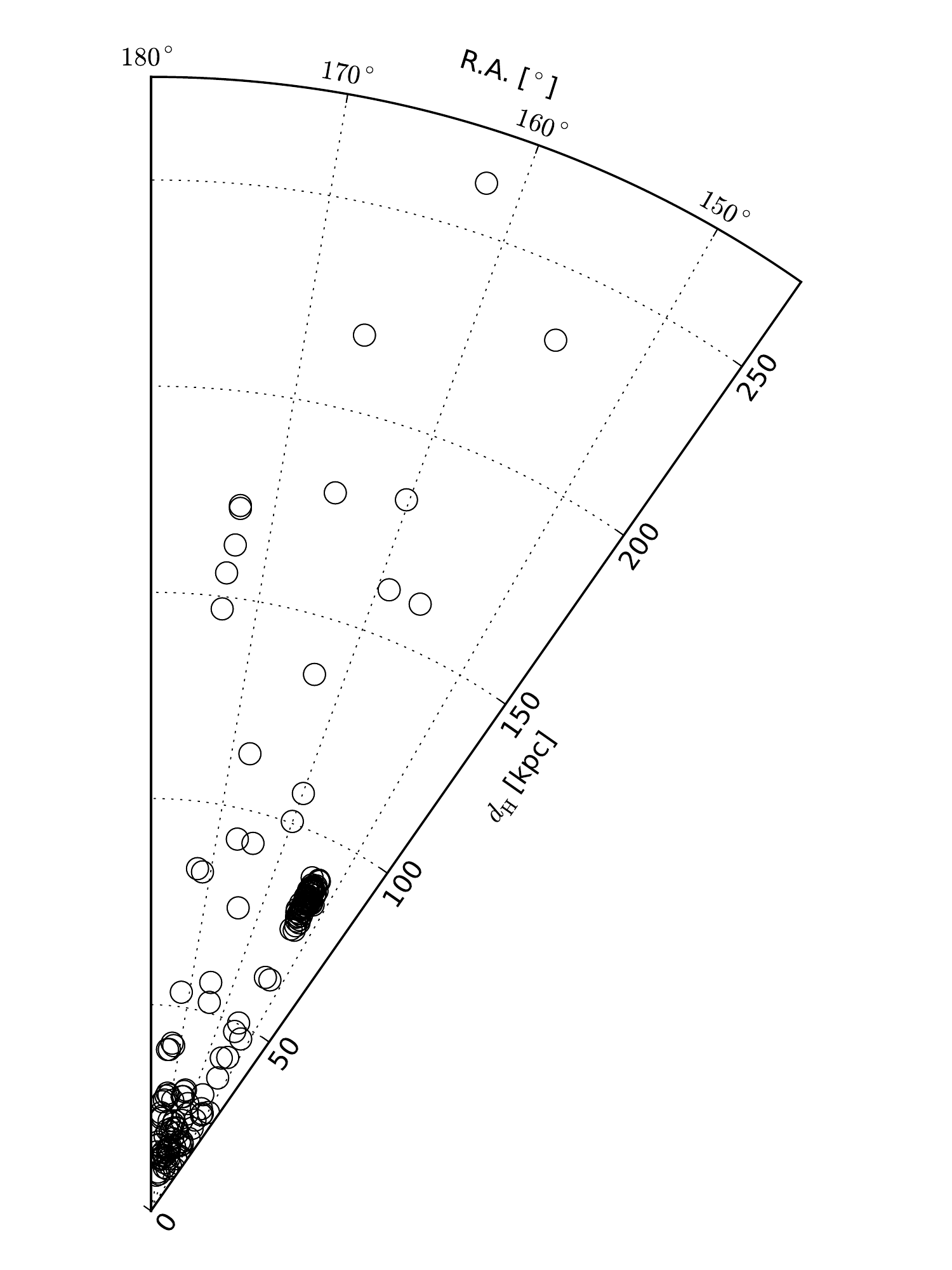}
\caption{Radial plot of heliocentric distances as a function of right ascension (R.A.). The overdensity located at $\sim 80$\,kpc corresponds to the Sextans dSph galaxy. Sixteen RRLs have distances $>100$ kpc.}
\label{fig:radial}
\end{figure}

RR Lyrae stars are known to follow a period-luminosity-metallicity (PLZ) relationship that makes them useful as distance indicators. We used \citet{Sesar2017a} PLZ relations to compute individual values of $M_g$, and subsequently heliocentric distances through distance modulus. For deriving the absolute magnitudes, we used the values from Table~1 in \citet{Sesar2017a} and assumed [Fe/H]$=-1.5$ as a representative value of the metallicity of the Galactic halo, given our lack of individual metallicities. 
It is worth noticing that the PLZ relationship is valid for RRc only when their periods are ``fundamentalized''. For the RRc in our sample, we used the periods given by

\begin{equation}
\log(P_F) =\log(P) +0.128 
\label{eq:logPf}
\end{equation}
\noindent where $P_F$ is the fundamentalized period and $P$ is the pulsational period of the RRc stars \citep{Catelan2009}. However, due to a discrepancy between the distances of RRab and RRc in one of the dwarf galaxies we study in this work, we applied an additional correction to the entire sample of RRc in our catalog (see Section 4).

The dependence of the absolute magnitude on metallicity is weak.
Considering a mean metallicity offset of $\pm0.5$\,dex from [Fe/H]$=-1.5$\,dex for the entire sample would lead to fractional difference in heliocentric distances of $1.8\%$, being these differences as high as $4$\,kpc (for the faintest RRL). If the metallicity offset is $\pm1.0$\,dex instead, these values vary $\sim 3.6\%$ and with distances up to $8$\,kpc. Hence, the lack of individual metallicities should not introduce large discrepancies in our distance estimates.
In Table~\ref{tab:distantRRL} the distances obtained by this method are shown for the faintest (and hence farthest) candidates. Distances obtained for the remaining sample are reported in Tables~\ref{tab:sextans} and~\ref{tab:tab_general} in the Appendix section. The uncertainties in the distance estimates include the propagation of the errors associated to not only the photometric errors but also the errors associated with $M_g$.
Figure \ref{fig:radial} presents a radial plot for the entire list. 
The HiTS RRLs are located between $9$ and $261$\,kpc from the Sun. The Figure shows there are a significant number of stars at heliocentric distances $>100$\,kpc, which we discuss in \S5. It is also significant the overdensity of stars near $d_{\rm H}\sim 80$\,kpc, which corresponds to RRLs in the Sextans dSph galaxy (see also Figure~\ref{fig:colordistancias}) that are discussed in the next section.

\section{Sextans dSph RRL population}

\begin{figure}
\includegraphics[angle=0,scale=.40]{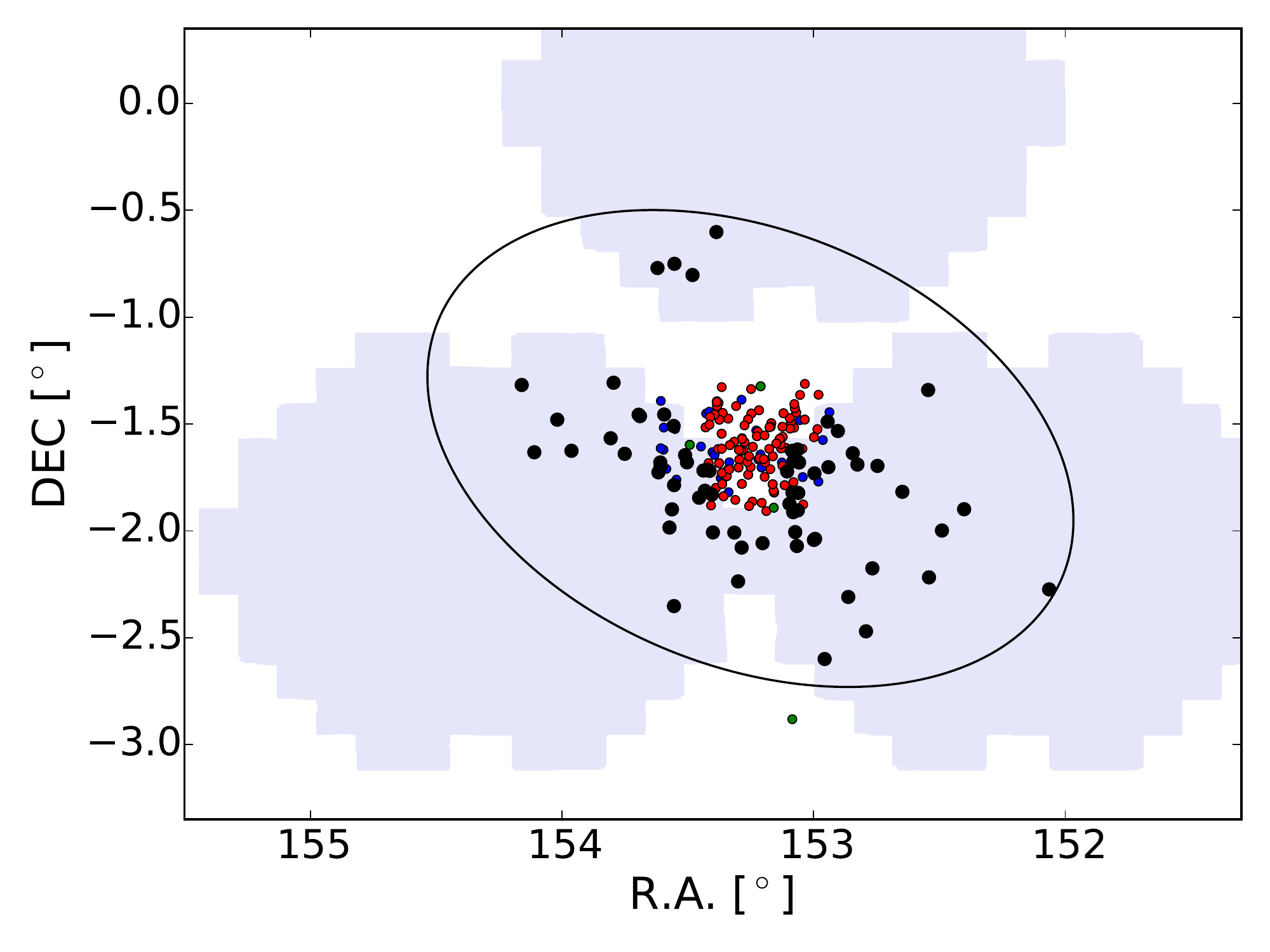}
\caption{Positions in the sky of RRLs found by this work near the Sextans dSph galaxy (black dots). These RRLs cover a range of magnitudes from $20.03$ to $20.57$ in the $g-$band. Known RRLs from \citet{Amigo2012} and \citet{Zinn14}, are shown as red and green dots, respectively. In addition, variable star candidates at the level of the Horizontal Branch ($19.0 < V < 21.6$) from \citet{Lee03} are plotted with blue dots.
The HiTS footprint is plotted in grey in the background. An ellipse centered in (RA,DEC) = $(153.2512^o, -1.6147^o)$ \citep{Irwin1990} is shown as a reference of the tidal radius ($r_t = 83.2'$), orientation ($\theta = 56.7^{\circ}$) and ellipticity ($\epsilon = 0.29$) of the dSph as determined by \citet{Roderick2016}.}
\label{fig:sextans}
\end{figure}

\begin{table*}\scriptsize
\caption{Most distant RR Lyrae stars ($d_{\rm H}>90$\,kpc). This table includes the main properties of the sample such as their distances, types, and number of observations (N). }
\label{tab:distantRRL}
\begin{center}
\begin{tabular}{|c c c c c c c|}
\hline
\hline
	ID  &  R.A.  &  DEC  & $\left<g\right>$ & $d_{\rm H}$  &  Type  &  N\\
	 &  (deg)  &  (deg)  &  & (kpc)  &   &  \\
\hline 
  HiTS105754-002603 & 164.47577 & -0.43403 & 20.5 & $92.5\pm6.8$ & c & 20\\
  HiTS110739+012813 & 166.91383 & 1.47037 & 20.4 & $92.6\pm6.6$ & c & 20\\
  HiTS104009-063304 & 160.03895 & -6.55105 & 20.8 & $100.5\pm4.0$ & ab & 21\\
  HiTS103943-021726 & 159.93119 & -2.29061 & 20.9 & $107.8\pm4.5$ & ab & 21\\
  HiTS111106-041718 & 167.77512 & -4.28834 & 21.2 & $113.4\pm8.8$ & c & 22\\
  HiTS105209-043942 & 163.03718 & -4.66174 & 21.5 & $136.0\pm5.7$ & ab & 20\\
  HiTS113259-003404 & 173.24674 & -0.56770 & 21.5 & $147.0\pm6.9$ & ab & 18\\
  HiTS113256-003329 & 173.23270 & -0.55818 & 21.5 & $155.8\pm7.5$ & ab & 19\\
  HiTS102414-095518 & 156.05905 & -9.92180 & 21.7 & $161.0\pm8.0$ & ab & 21\\
  HiTS103601-015451 & 159.00456 & -1.91422 & 21.7 & $161.3\pm12.9$ & c & 21\\
  HiTS113107+021302 & 172.77796 & 2.21734 & 21.7 & $162.8\pm8.6$ & ab & 21\\
  HiTS113057+021331 & 172.73946 & 2.22514 & 21.8 & $171.7\pm9.4$ & ab & 20\\
  HiTS113105+021319 & 172.76936 & 2.22200 & 21.8 & $172.4\pm9.4$ & ab & 20\\
  HiTS110222-001624 & 165.59251 & -0.27337 & 22.1 & $179.8\pm10.3$ & ab & 19\\
  HiTS104054-042827 & 160.22661 & -4.47424 & 21.9 & $183.2\pm14.8$ & c & 20\\
  HiTS110510-022710 & 166.28982 & -2.45282 & 22.4 & $218.6\pm14.6$ & ab & 19\\
  HiTS102014-042354 & 155.05789 & -4.39843 & 22.5 & $232.9\pm21.9$ & c & 19\\
  HiTS104738+020627 & 161.90718 & 2.10746 & 22.8 & $262.2\pm24.3$ & c & 19\\
\hline
\end{tabular} 
\end{center}
\end{table*}

Figures \ref{fig:colordistancias} and~\ref{fig:radial} show an overdensity of stars centered at RA $\simeq 153.3^\circ$, DEC $\simeq-1.7^\circ$ and $d_{\rm H}\simeq 80$\,kpc. The position of these stars, both in equatorial coordinates and distance, matches the location of the Sextans dwarf spheroidal galaxy
\citep[R.A. = $153.2512^\circ$, DEC = $-1.6147^\circ$, $d_{\rm H} = 86\pm 4$\,kpc,][]{McCon12}.
Sextans is a Milky Way satellite discovered by \citet{Irwin1990} with an absolute
magnitude of $M_V = -9.3$. The system is characterized by a relatively old and metal poor population (age=$12$\,Gyr; [Fe/H]=$-1.9$), as described by \citet{Mateo91} and \citet{Kirby2011}, respectively. The Sextans dwarf is a relatively extended satellite with a half-light radius $r_h$ of $695$\,pc \citep{Irwin95, Roderick2016} and a surface brightness of $\sim 28$ mag\,arcsec$^{-2}$, typical of Local Group dSph galaxies. A recent study of Sextans carried out by \citet{Roderick2016} analysed its structural parameters using wide-field photometric data in order to investigate its kinematics and stellar structures. According to their results, the dSph has a halo-like substructure extended up to $82\arcmin$ from its center, with several overdensities detected at statistically high significance levels. 

From our sample, $65$ of the RRL candidates are located within $1.75$\,degrees of Sextans, and between heliocentric distances of $76$ and $90$\,kpc ($48$ ab's, $17$ c's). Their distribution, as well as the position of the HiTS fields are shown in Figure~\ref{fig:sextans} and Figure~\ref{fig:colordistancias}, in different scales. Figure~\ref{fig:sextans} also shows an ellipse marking the tidal radius from a King profile of the galaxy \citep{Roderick2016} and known RRLs in the galaxy from the literature. As can be seen in the figure, the HiTS fields do not cover the center of Sextans but only the outskirts of the dwarf. However, none of the previous surveys of RRLs in this galaxy were able to cover the full extension of Sextans. Thus, HiTS is providing for the first time information on the outermost RRLs in Sextans. The distribution of the RRLs along the fields does not show any particular kind of structure or shape and all are contained within the galaxy's known King limiting radius. 

The distances for the Sextans dSph RRLs were re-calculated since values for the metallicity of the galaxy are available in the literature, and they are not necessarily close to our assumption for the Halo ([Fe/H]$=-1.5$). For the Sextans RRLs, we adopted [Fe/H]$=-1.93\pm0.01$ from \citet{Kirby2011}.

Based on our subsample of RRLs associated to Sextans (both $ab$ and $c-$types), in principle we estimated a mean heliocentric distance to the satellite of $81.4\pm 5.7$\,kpc. We noted, however, a clear offset between the mean distance obtained with only RRab and only RRc ($84.2$ and $74.5$\,kpc, respectively). This discrepancy (13\%) may be due to the different behavior of the period-luminosity relations for these stars \citep[see for example][]{Vivas17}, and we are using PLZ relations that are exclusive for RRab \citep{Sesar2017a}.  
As we do not expect the different populations to be located at different distances, we corrected the the distances of the RRc from our entire sample by this number, including all type c stars outside Sextans.
By doing this, the re-derived mean heliocentric distance of the dSph is $84.2\pm3.3$\,kpc, which is in agreement with the distance of $86\pm4$ kpc obtained by \citep{Mateo1995} based as well on RRLs.

Knowing the distance we can estimate the physical distances of our Sextans RRLs to the dwarf's center. We find stars out to $1.9$\,kpc ($1.30^\circ$) which is in agreement with the results of \citet{Roderick2016}, who found Sextans halo overdensities to be as far as $2$\,kpc from the center.

RRLs have been used as a mean to detect extra-tidal material around satellite galaxies and globular clusters \citep[eg.][]{Fernandez15,Garling17}.
From Figure~\ref{fig:sextans} it is possible to claim that there is no clear evidence of extra-tidal RRLs based on our sample. Extending the search for RRLs in Sextans to the contiguous HiTS fields does not change this statement, even if a significantly larger radius is considered \citep[e.g. $r_t=90'$;][]{Irwin1990}.
One of the RRLs from the LSQ survey lies outside \citeauthor{Roderick2016}'s tidal radius (see Figure~\ref{fig:sextans}), but inside the estimate by \citet{Irwin1990}. Thus, it is not strong evidence of extra tidal material.
Regarding hypothetical internal substructures in the galaxy, it is not possible to measure any given the mean uncertainty of the distance of our sample of Sextans' RRLs ($4.1$\,kpc), the standard deviation of the distribution ($3.3$\,kpc), and its size \citep[$\sim 700$\,pc; ][]{Irwin95, Roderick2016}. 

In \citet{Amigo2012}, a list of $114$ RRLs covering the inner regions of Sextans is presented, including $37$ identified earlier by \citet{Mateo1995}. This catalog contains a large number of new discoveries (ranging from $V=20.06$ to $V=20.50$), and recovers as RRLs most of the variable star candidates presented in \citet{Lee03}. 
Since \citet{Amigo2012} covers the central region of the dwarf, the overlap with HiTS' fields is not significant. However, there are $16$ stars in common between HiTS and that work, with highly similar periods ($|\Delta P|\sim2\times 10^{-4}$) and same classifications. In addition, \citet{Zinn14} recognized seven RRLs in the galaxy, four of which are not contained in the catalog of \citet{Amigo2012}. Thus, there are $46$ new RRLs discovered by HiTS in Sextans, seven of which are not flagged as RRLs by \citet{Lee03}. This brings the total number of RRLs in this galaxy to $165$. 

Folded light-curves for the HiTS Sextans RRLs sample are shown in the appendix section (Figure~\ref{fig:lcSextans1} to Figure~\ref{fig:lcSextans3}). Table~\ref{tab:sextans} contains the information of these stars. The mean period of the RRab is $0.63$\,d which is close to the nominal Oo II group. However, the distribution of the $65$ RRLs in Sextans in the Period-Amplitude diagram (Figure~\ref{fig:peramp}) clearly shows that the stars do not follow a single sequence of Oo groups. Thus, HiTS confirms the Oo-intermediate nature of this dSph which was already established by \citet{Mateo1995} based in $34$ stars in the inner part of the galaxy. Following the Milky Way satellites (except the massive galaxies of Sagittarius and the Magellanic Clouds) our extended sample confirms that Sextans do not contain High Amplitude Short Period (HASP) stars, which have been interpreted by \citet{Fiorentino2015} as coming from populations more metal rich than [Fe/H]$\sim -1.5$.

\section{RR Lyrae stars beyond 90 kpc}

\begin{figure*}[!t]
\center
\includegraphics[height=0.8\textheight]{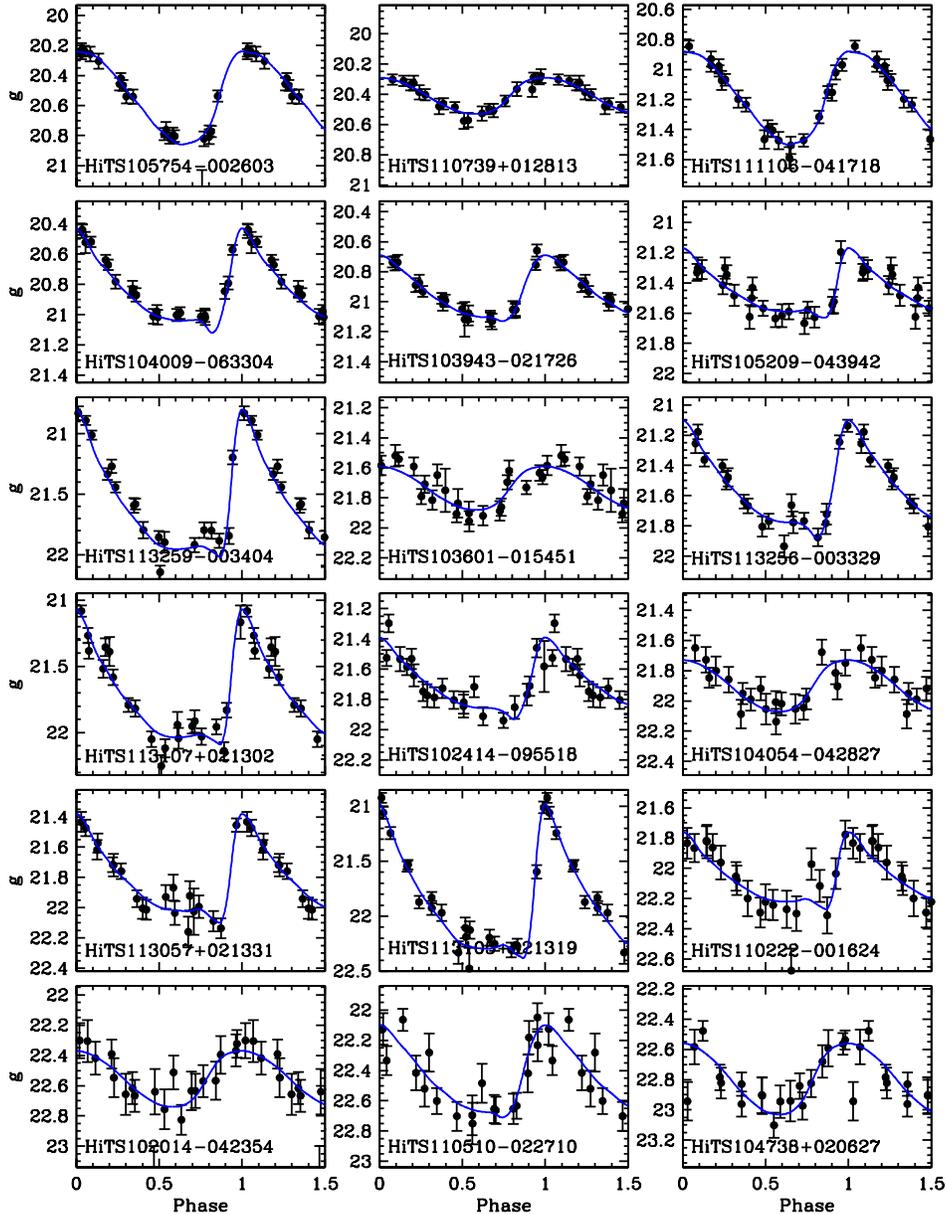}
\caption{Folded light curves of the distant RR Lyrae stars listed in Table~\ref{tab:distantRRL}.}
\label{fig:lcdistant}
\end{figure*}

We find $18$ distant RRLs within our list of candidates that are located at $d_{\rm H} > 90$\,kpc (hereafter distant RRLs); this corresponds to mean magnitudes $\left<g\right> \geq 20.5$.
Two recent works, \citet{Dra13_100kpc} and \citet{Sesar2017b}, have reported some of the most distant Galactic RRLs known to date, located at $\sim120$\,kpc and $\sim130$\,kpc, respectively.
Among our distant RRLs sample there are $13$ with $d_{\rm H} > 130$\,kpc; i.e., beyond the most distant previously known MW RRLs.
The distant RRLs group span a range from $\sim 92$\,kpc ($\left<g\right>\sim 20.5$) to beyond $200$\,kpc (up to $\left<g\right>\sim 22.8$). In Table~\ref{tab:distantRRL} the main properties of these distant stars are presented (full info is presented in Table~\ref{tab:tab_general} in the Appendix section), and Figure~\ref{fig:lcdistant} shows the folded light curves for this group.

Regarding their classification, $11$ RRL ($61\%$) from this faint subsample correspond to RRab, while seven were classified as RRc (Table~\ref{tab:distantRRL}), roughly consistent with the observed ratio in the rest of the sample. We note that three of the four most distant RRLs are $c-$type. Given their lower intrinsic amplitude and the large photometric uncertainties of the individual observations near the detection limit of our survey, we regard their classification as tentative; the most distant RRab, which has a secure classification, is at $d_{\rm H} = 219 \pm 15 $\,kpc. In the Period-Amplitude diagram (Figure~\ref{fig:peramp}, top panel) these stars are mostly located toward the locus of the Oo II group, although a large dispersion is observed. However, the distribution looks significantly different compared with the one from nearby stars ($d_{\rm H}<90$\,kpc). The mean period of the sample of $11$ distant RRab stars is $0.668 \pm 0.05$\,d, which is significantly different to the mean period of the RRab with $d_{\rm H}<90$\,kpc, $0.600 \pm 0.09$\,d. The mean period of the distant RRLs is remarkably similar to the mean period of the ensemble of RRLs in the UFD satellites of the Milky Way, $0.667$\,d \citep{Viv16}, suggesting that these galaxies may be the main contributors of the outermost regions of the halo. As Sextans, the distant RRLs sample does not contain HASP stars.

In Figure~\ref{fig:colordistancias} we showed the spatial distribution of the distant RRLs, color coded by their heliocentric distance. The distant RRLs look randomly distributed in the sky except for two compact groups, with two and three RRLs  respectively, whose stars are very close together in the sky. We discuss those groups in the next sub-section.

\subsection{Leo~IV and Leo~V}
Among the sample of distant RRLs, we found two distant groups of closely spaced RRL both in angular separation and heliocentric distances. In these cases, they coincide with the position on the sky and distance of the ultra-faint galaxies Leo~IV and Leo~V. 
In \citet{Medina2017} we reported the discovery of the RRLs in Leo~V since that galaxy was not explored before for variability. \citet{Medina2017} also argues the importance of groups of distant RRLs as a mean to discover previously unknown satellites or substructure in the halo. Since the properties of the RRLs in Leo~IV and Leo~V were already discussed in \citet{Medina2017}, here we only present a summary and update the distance estimates.

Out of the three known RRLs in Leo~IV \citep{Moretti2009}, we detected two stars, HiTS113256-003329 and HiTS113259-003404, both RRab. The non-detection of the third RRLs (called V3 by \citealt{Moretti2009}) is consistent with the detection efficiency expected for RRLs of these magnitudes, according to our results (Section 3). 
Three RRLs were identified as members of Leo~V, HiTS113057+021331, HiTS113105+021319, and HiTS113107+021302, all being new discoveries.

In \citet{Medina2017}, the distances of the Leo~IV and Leo~V RRLs were anchored to the known distance to Leo~IV \citep{Moretti2009}.
Here, we obtain revised distances for those RRLs since metallicities for both systems are available in the literature and we can use them in the PLZ relation given by \citet{Sesar2017a}, which was not available at the time of writing \citet{Medina2017}.
Using [Fe/H]$=-2.31$ for Leo~IV \citep{Simon2007}, we get a mean heliocentric distance of $151.4 \pm 4.4$\,kpc. This number is in agreement with the estimation made by \citet{Moretti2009} ($154 \pm 5$\,kpc).
For Leo~V, assuming [Fe/H]$=-2.48$ \citep{Collins2016} gives us a mean distance of $169.0 \pm 4.4$\,kpc, which is still consistent with our earlier results.

\section{Space density distribution}

The density profile of halo tracers contains important clues of the accretion history of the Milky Way \citep{BJ05,Cooper2013}.
Since RRLs are excellent distance indicators they are particularly useful for the construction of number density profiles $\rho(R)$, i.e., the variation in the number of RRLs per unit volume ($\#$ kpc$^{-3}$) where $R$ is the distance to the center of the Galaxy. Recent studies have worked with catalogs of RRLs sufficiently large that allowed them to study the variation of the number density with direction in the sky \citep[see for example][]{Zinn14}.
In our case, the sample is made up of only $\sim100$ RRLs (excluding stars from known dSph galaxies) spread over an area of $\sim120$\,deg$^{2}$. 
Therefore, working with sub-samples located in different directions would not be statistically meaningful.
For this reason, we use the data from all $40$ fields to build 
a single number density profile. 
Since HiTS' fields span from $39$ to $60$\,deg in Galactic latitude, and from $236$ to $269$\,deg in longitude, this can be considered a single line of sight in the Galaxy. For the selection of the distance bins and the number density calculation, the distance from the Galactic center $R_{\rm gc}$ was calculated for each star in our sample. $R_{\rm gc}$ can be obtained from

\begin{equation}
R_{\rm GC}^2=(R_\odot - d_{\rm H} \cos{b} \cos{l})^2 + d_{\rm H}^2 \sin^2{b} + d_{\rm H}^2 \cos^2{b} \sin^2{l},
\label{eq:Rgc}
\end{equation}

\noindent where $d_{\rm H}$ is the heliocentric distance, $b$ and $l$ are the Galactic latitude and longitude of each star, respectively, and $R_\odot$ is the distance from the sun to the Galactic center. For this work we assumed $R_\odot$ to be $8$\,kpc. The bins were created evenly-spaced on a log scale.

We are interested in placing the Milky Way into context with external galaxies and model stellar halos. Given that in typical data sets, small overdensities of 2-3 stars, such as what we found in Leo~IV and Leo~V, may not have been identified with known galaxies, we first performed density profile fits including the RRLs in the UFD, but leaving out the RRLs from the larger Sextans dSph. We repeated the fits with the Leo~IV/V stars removed, in order to test the effect of these ultra-faint dwarfs on the profile; the fits are virtually unchanged by the inclusion/exclusion of the Leo~IV/V RRLs.

Two halo models were used to fit the data: a spherical (sph) halo model and an ellipsoidal (ell) one with a flattening parameter $q=0.7$ adopted from \citet{Ses11}. The adopted model was $\rho(R) = \rho_\odot(R/R_\odot)^n$, where $\rho_\odot$ is the local (solar circle) number density of RRLs. The fit was done using the logarithmic form of the previous equation:

\begin{figure*}[!t]
\begin{center}
\includegraphics[width=0.65\textwidth]{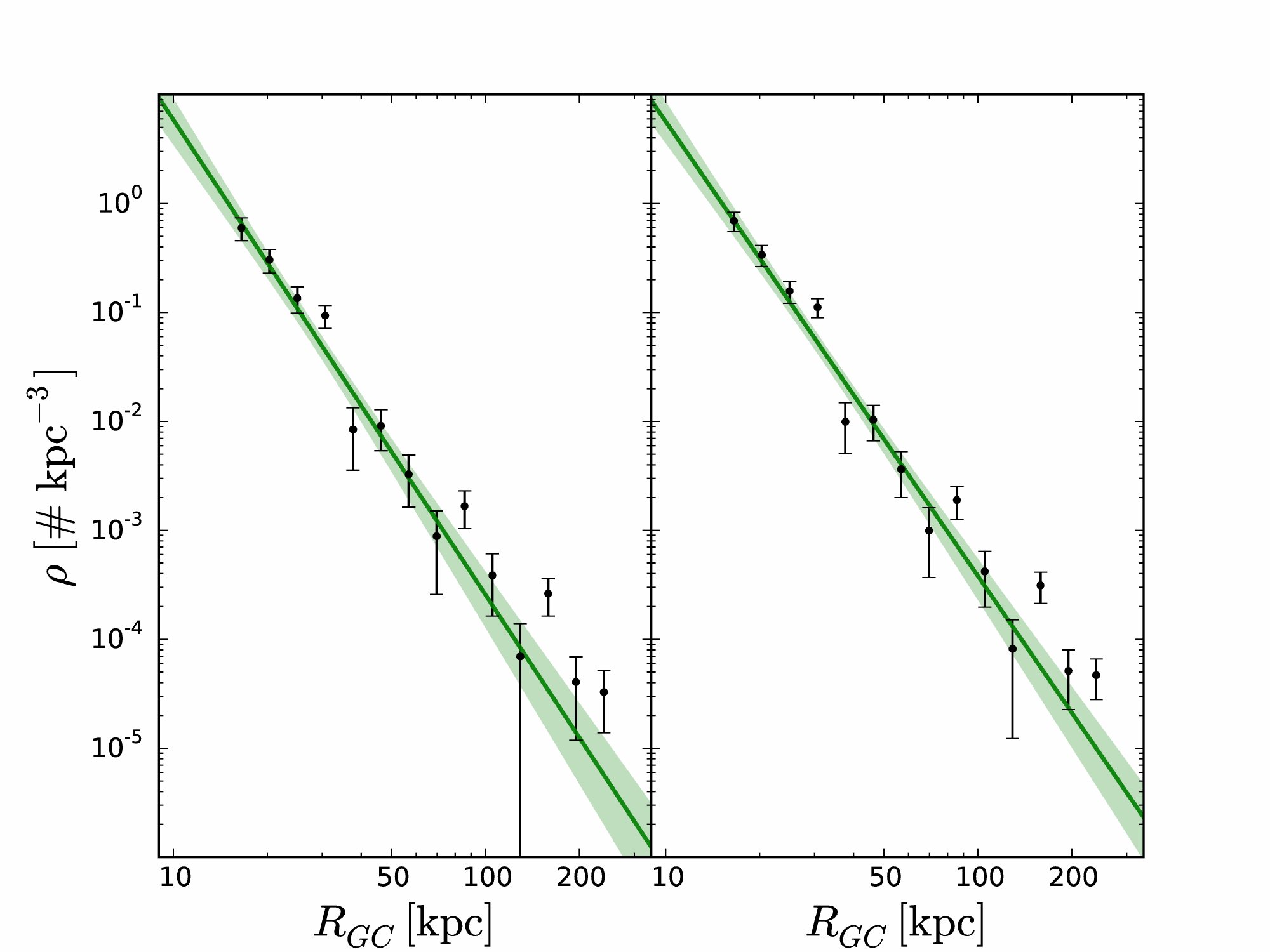}
\caption{Number density versus Galactocentric distance $R_{\rm GC}$, excluding RR Lyrae stars from the Sextans dwarf galaxy, assuming a spherical halo. Density profiles built without and with considering our detection efficiency (see Figure~\ref{fig:completitud}) are shown in the {\it left} and {\it right} panels, respectively. We fit a simple power-law model to the corrected data; the corresponding fit parameters are listed in Table~\ref{tab:parameters}, and the fits overlaid as solid lines in both panels. 
The shaded regions show the $3\sigma$ confidence levels determined via MCMC methods.
}
\label{fig:fitRgc}
\end{center}
\end{figure*}

\begin{figure*}[!t]
\begin{center}
\includegraphics[width=0.65\textwidth]{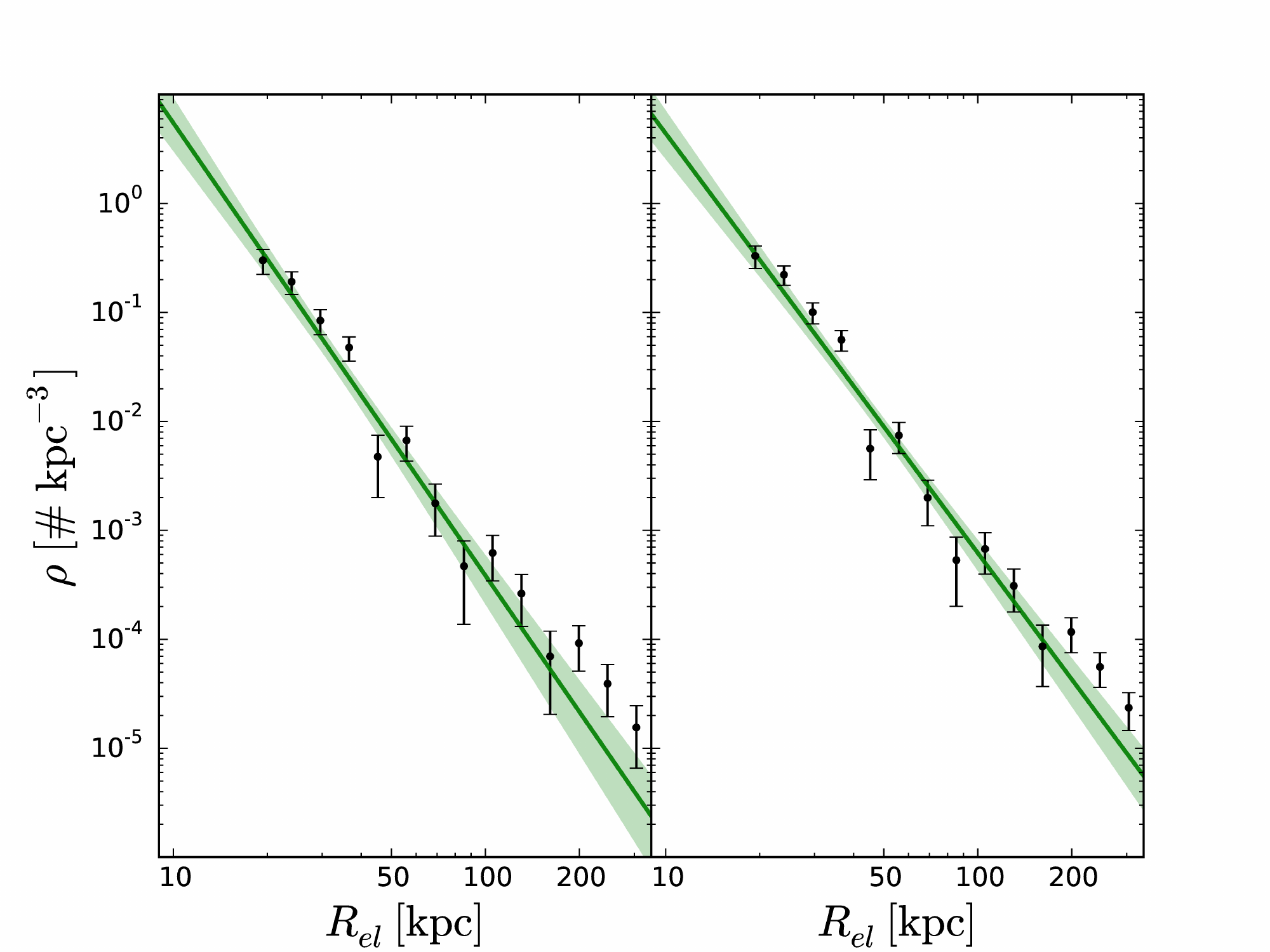}
\caption{Number density as a function of radius in an elliptical halo, $R_{\rm el}$, assuming an oblate halo with $q = 0.7$. As in Figure~\ref{fig:fitRgc}, we exclude RRLs from the Sextans dSph. Density profiles built without and with considering our detection efficiency (see Figure~\ref{fig:completitud}) are shown in the {\it left} and {\it right} panels, respectively. 
We fit a simple power-law model to the data at uncorrected $R_{\rm el} < 145$\,kpc; the corresponding fit parameters are listed in Table~\ref{tab:parameters}, and the fits overlaid as solid lines in both panels. 
The shaded regions show the $3\sigma$ confidence levels determined via MCMC methods.}
\label{fig:fitRel}
\end{center}
\end{figure*}

\begin{equation}
\log{(\rho (R))}=A+n\ \log{(R/R_\odot)}
\label{eq:logarithm}
\end{equation}

\noindent where $A=\log{(\rho_{\odot})}$. We used the Markov Chain Monte Carlo routine \it{emcee} \rm \citep{Foreman2013} to find distributions for the parameters of this model of a simple power law (SPL), $A$ and $n$.

Although our main interest is to study the behaviour of the density profile at large distances from the Sun, we also explored a model with a broken power law which has been observed in multiple works \citep{Saha85, Wat09, Deas11, Ses11} at $\sim 25$ kpc. For the broken power law model (BPL), the break radius was considered a free parameter, as well as the inner and outer slopes. In the case of the BPL profile, the equations used are:

\begin{equation}
\begin{array}{ccc}
\log{(\rho (R))}=A1+n_1\ \log{(R/R_\odot)}\\
\log{(\rho (R))}=A2+n_2\ \log{(R/R_\odot)}\\
A1+n_1\ \log{(R_{break}/R_\odot)}=A2+n_2\ \log{(R_{break}/R_\odot)}
\end{array}
\label{eq:logarithm_2}
\end{equation}

\noindent The values of $A$ were constrained to within $0.20$ and $1.75$, and $-5$ and $5$ for $A2$ (after the break).
The slope was constrained to be between $-7$ and $-1$ for the SPL. For the BPL, the constraining values were set to between $-10$ and $5$, and $-14$ and $15$ for the inner and outer slopes, respectively. The break radius was constrained to within $15$ and $70$\,kpc.
An initial guess for the value of each parameter was given, based on the result of following a non-linear least squares methodology (using the Levenberg-Marquardt algorithm from \it{scipy}).\rm

The \textit{left}/\textit{right}\rm \ panels of Figure~\ref{fig:fitRgc} and Figure~\ref{fig:fitRel} show the fitted models to the data, with sph and ell halos respectively, under different considerations. In the \textit{left} panels we fit the model to the RRLs within $145$\,kpc, which corresponds to the upper limit based on completeness considerations (we are $\sim85\%$ complete down to $g=21.5$). In this case, the more distant ($R >145$\,kpc) bins are shown for comparison with extrapolations of the fits to large radii.
In the \textit{right} panels, we show fits to the entire sample, after correcting for detection efficiency as calculated in Figure~\ref{fig:completitud}.

\begin{figure}
\begin{center}
\includegraphics[scale=0.5]{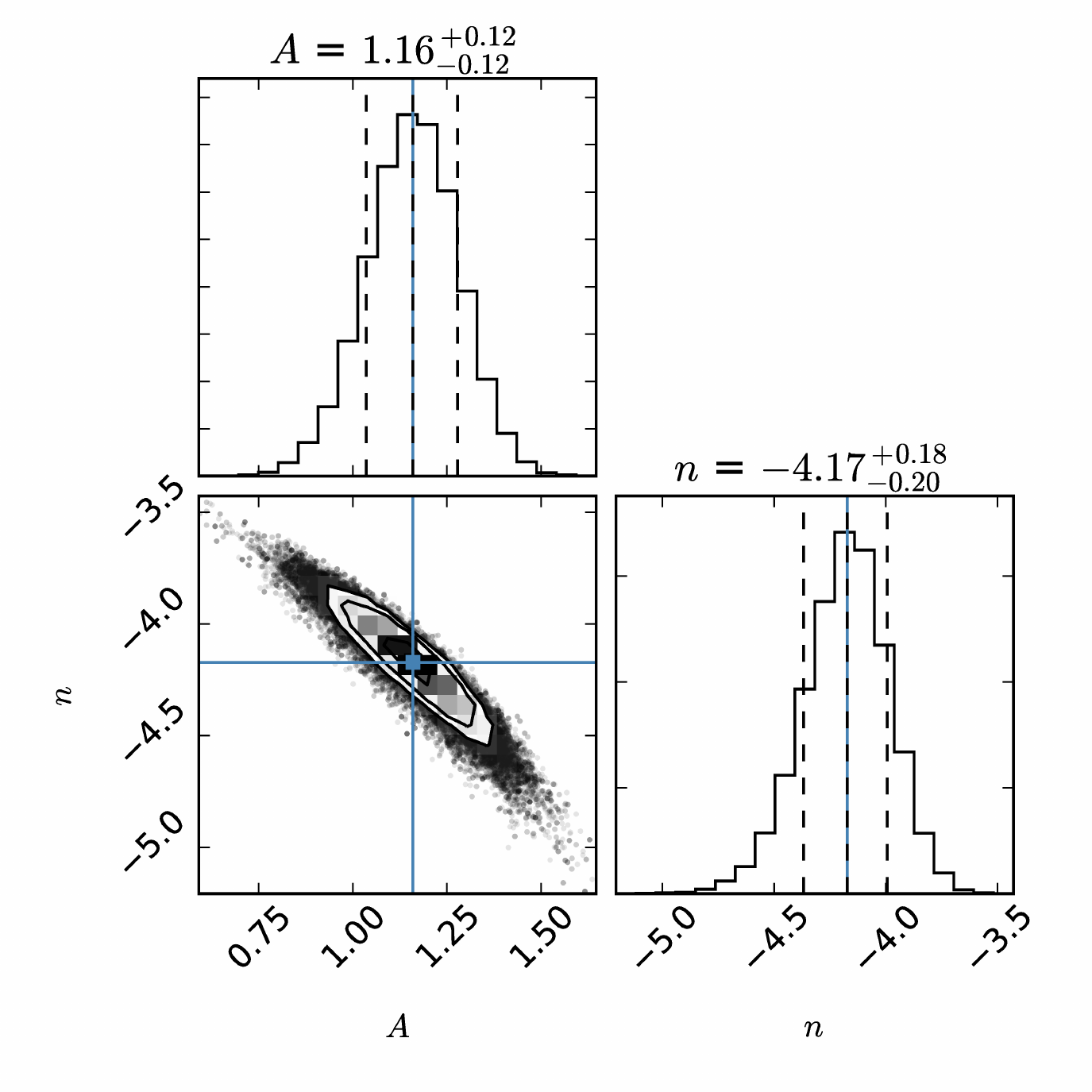}
\caption{Corner plot showing the posterior joint probability distribution of the parameters of a simple power law number density profile, for a spherical Halo model. The parameters used for this model are the logarithm form of the number density in the solar neighborhood ($A$) and the power law index of the model ($n$).}
\label{fig:simplePL}
\end{center}
\end{figure}

In both cases the first bin, which includes the closest RRL, was removed from the plots and the analysis because these bins suffer incompleteness due to saturation of bright stars, which was not considered in our detection efficiency calculations. The corner plot with the resulting posterior distributions from our fits are shown in (Figure~\ref{fig:simplePL}). Tables~\ref{tab:parameters} and ~\ref{tab:parameters_noLeo} shows the results of the fitting for the different models assumed and with the samples with/without the RRLs in Leo IV and V, respectively.

It is important to notice that, according to Equation ~\ref{eq:Rgc}, the Galactic latitude and longitude are required to determine $R_{\rm{GC}}$ from $d_{\rm H}$. Given the wide area covered by the survey, it is not possible to transform the detection efficiency per apparent magnitude bin to a unique Galactocentric distance and subsequently to unique ellipsoidal distance. 
If the mean latitude and longitude of the area observed by the survey is used ($b\sim 46^\circ$, $l\sim 250^\circ$), the difference in distance, between heliocentric and Galactocentric, is of up to $5$\,kpc. 
We consider the detection efficiency correction still valid in this case, since the size of the bins is in general $>5$\,kpc. In the case of the ellipsoidal distances, this distance difference reaches up to $70$\,kpc, which makes the correction less accurate for the more distant bins. The \textit{left}/\textit{right}\rm \ panel of the plots in Figure~\ref{fig:fitRgc}/Figure~\ref{fig:fitRel} are provided only to show the approximate effect to the density profile if the detection efficiency is or is not considered. 
For this reason, the best-fit parameters in Tables~\ref{tab:parameters} and ~\ref{tab:parameters_noLeo} and the posterior analysis are only given for the corrected/uncorrected data, for the spheroidal/ellipsoidal model.

\begin{table*}\scriptsize
\caption{Parameters for the different power law models described in section 6, with the Leo~IV and Leo~V RRLs.}
\label{tab:parameters}
\begin{center}
\begin{tabular}{|c c c c c c c|}
\hline
\hline
	Model  &  A1  &  A2  &  n1  &  n2 & $R_{b}$ & $\chi^2_\nu$\\
	 &    &       &    &  (kpc)  & (kpc) & \\
\hline
sph SPL    &  $1.16^{+0.12}_{-0.12}$  &  --  &  $-4.17^{+0.18}_{-0.20}$  &  --  &  -- & $2.657$  \\  
sph BPL    &  $0.91^{+0.56}_{-0.70}$  &  $1.20^{+0.17}_{-0.25}$  &  $-3.50^{+1.98}_{-1.46}$  &  $-4.22^{+0.32}_{-0.26}$  & $19.52^{+9.90}_{-3.27}$ &  $3.204$   \\

& & & & & &
  \\\ 
ell SPL    &  $1.14^{+0.15}_{-0.16}$  &  --  &  $-4.16^{+0.23}_{-0.25}$   &  --  & --  &  $3.184$  \\  
ell BPL    &  $0.42^{+0.75}_{-0.74}$  &  $1.30^{+0.37}_{-0.24}$  &  $-2.23^{+1.77}_{-1.99}$  &  $-4.37^{+0.34}_{-0.50}$  &  $22.46^{+7.10}_{-5.50}$ &  $3.962$   \\

\hline
\end{tabular} 
\end{center}
\end{table*}

\begin{table*}\scriptsize
\caption{Parameters for the different power law models described in section 6, without Leo~IV and Leo~V.}
\label{tab:parameters_noLeo}
\begin{center}
\begin{tabular}{|c c c c c c c|}
\hline
\hline
	Model  &  A1  &  A2  &  n1  &  n2 & $R_{b}$ & $\chi^2_\nu$\\
	 &    &       &    &  (kpc)  & (kpc) & \\
\hline
sph SPL    &  $1.18^{+0.11}_{-0.12}$  &  --  &  $-4.22^{+0.18}_{-0.19}$  &  --  &  -- & $2.132$  \\
sph BPL    &  $0.75^{+0.66}_{-0.67}$  &  $1.24^{+0.16}_{-0.19}$  &  $-3.01^{+2.11}_{-1.77}$  &  $-4.30^{+0.28}_{-0.25}$  & $18.85^{+7.76}_{-2.67}$ &  $2.584$   \\

& & & & & &
  \\\ 
  
ell SPL    &  $1.14^{+0.16}_{-0.16}$  &  --  &  $-4.15^{+0.23}_{-0.25}$   &  --  & --  &  $2.497$  \\ 
ell BPL    &  $0.43^{+0.78}_{-0.88}$  &  $1.29^{+0.33}_{-0.23}$  &  $-2.29^{+2.18}_{-2.02}$  &  $-4.37^{+0.33}_{-0.45}$  &  $22.04^{+6.43}_{-5.26}$ &  $3.131$   \\
\hline
\end{tabular} 
\end{center}
\end{table*}

\subsection{Spherical Halo}

In this case $R^2=R_{\rm{GC}}^2 = X^2+Y^2+Z^2$, where $X$ and $Y$ are the cartesian coordinates in the Galactic plane, and $Z$ is the axis perpendicular to it. These values are given by:

\begin{equation}
\begin{array}{lcl}
x = R_\odot -d_{\rm H}\ \cos{b}\ \cos{l}\\
y = d_{\rm H}\ \cos{b}\ \sin{l}\\
z = d_{\rm H}\ \sin{b}
\end{array}
\label{eq:xyz}
\end{equation}

Figure~\ref{fig:fitRgc} shows the density profiles as described above. 
Table~\ref{tab:parameters} summarizes the parameters found for the spherical halo SPL and BPL, when the data is corrected and not corrected by detection efficiency, respectively. 
Based on these results, the best fit for the data corresponds to the SPL model. From Figure~\ref{fig:fitRgc} one can infer that the presence of the break at $\sim 20$\,kpc is mostly due to the relatively-high density of the point at $R_{\rm GC}\sim 26$\,kpc. 
The larger uncertainties in the determination of the BPL parameters, compared with the ones for the SPL, support the low probability of that model. The lack of stars at short distances in our sample may be part of the reason in finding a good fit to this model.
For completeness, we also fit a broken power law with two breaks (double broken power law; DBPL)
to account for a potential different behavior of the profile at large radii, ruled by the interaction of the Milky Way halo with farther large massive galaxies (mostly M31), for example. Table~\ref{tab:parameters} does not include these parameters, given the large uncertainties associated and the difficulty to assess a physical meaning to this fit.

\subsection{Ellipsoidal Halo}

For an oblate halo with $q=0.7$, the semi-major axis of the ellipsoid is $R_{\rm el}^2=X^2+Y^2+(Z/0.7)^2$ which replaces $R$ in Equation~\ref{eq:logarithm}. The results of the fit are shown in Table~\ref{tab:parameters} and Figure~\ref{fig:fitRel}. As in the previous case, the best fit corresponds to the SPL model (based on $\chi^2_\nu$) but again, this may be due to the scarse bins at short distances in our dataset.
As in the case of the spherical halo model, the distant bins make the difference when looking for a change in the slope. 
The fits give $R_{\rm el}\sim 22$\,kpc as the most likely value for the break radius for the BPL, while for the DBPL these are $\sim25$\,kpc and $\sim80$\,kpc, with large uncertainties. For the latter, overfitting is much more likely to have happened than in the spherical halo model. 

\bigskip
Overall, the spherical halo model provides a better fit to the halo than the ellipsoidal model with $q=0.7$. The results of using samples that include/exclude the RRLs in the UFD Leo~IV and Leo~V are the same, within the erros, indicating that small galaxies do not affect the general behaviour of the density profile of the halo.

\subsection{Contribution from known substructures in the HiTS field of view}

The approach we have taken here best facilitates comparisons to model and external galaxy halos, where individual stellar streams and substructures may not be distinguished from the halo as a whole. However, we acknowledge that there are known MW substructures in the vicinity of the HiTS field of view. Here we briefly assess these substructures' contributions to the halo density profiles shown in Figures~\ref{fig:fitRgc} and \ref{fig:fitRel}.

The Orphan tidal stream \citep{Grillmair06_Orphan,Belokurov07} is a wide stellar stream that spans more than $100$ degrees \citep{Grillmair15_Orphan}, tracing a roughly north-south path across (and extend southward beyond) the SDSS footprint. The distance to the stream ranges from $\sim 55$\,kpc at its northernmost point, decreasing to $\sim 20$\,kpc as the Orphan extends below the southern edge of the SDSS footprint \citep{Sesar13}. We select Orphan RRLs candidates using the coordinate system defined by \citet{Newberg10} to align with the stream and at distances between $20 < d_{\rm H} < 32$\,kpc (consistent with the stream distances found in the region of the HiTS footprint by \citealt{Hendel17}). We find 12(7) RRLs within $5^\circ(3^\circ)$ of the stream center and satisfying this distance cut. \citet{Newberg10} showed that the Orphan stream is a $\sim10\%$ excess of MSTO stars relative to the nearby field density; thus, we estimate that $\sim1-2$ of the stars selected as Orphan candidates are actual members of the stream. Indeed, one of them (HiTS104924-023635) corresponds to ``RR49'' from \citet{Hendel17}, who measured a distance with Spitzer Space Telescope time-series data of $23.05\pm0.58$\,kpc. This compares favorably to our estimate of $d_{\rm H} = 20.92\pm0.75$\,kpc for the same star, especially given that if we instead adopted its metallicity of [Fe/H]$=-2.02$ as measured by \citet{Sesar13}, we would derive a slightly larger distance (by $\sim2\%$).

Another possible contributor to the halo stellar populations in the HiTS field of view is the Sagittarius (Sgr) tidal stream. Sgr is a prominent stellar overdensity over much of the sky (see review in \citealt{LawMajewski16}). Using the $N$-body model of \citet{Law10}, we select model Sgr stars within the footprint of our HiTS RRLs search, and that were stripped from the progenitor within the last $\sim5$\,Gyr (i.e., are part of the most recent leading debris tail). The model predicts that only the northeastern corner of the HiTS footprint should contain any Sgr debris. 
The mean distance of these model debris points is $\left<d\right> = 42$\,kpc, with a spread of $\sigma_{\rm d} \sim 6$\,kpc. There are nine HiTS RRLs between $31 < d_{\rm H} < 49$\,kpc, of which four are in the region occupied by Sgr model debris (i.e., $170^\circ < \rm{7RA} < 175^\circ$, $-3^\circ < \rm{Dec} < 3^\circ$). Interestingly, these four stars are tightly clumped in distance, with a mean of $\left<d_{\rm H}\right> = 40.0$\,kpc and standard deviation of $\sigma_{\rm d} \sim 0.7$\,kpc. Nonetheless, given that these four stars are $\sim16.5-20^\circ$ from the center of the Sgr stream (using the Sgr-aligned coordinate system of \citealt{Maj03}), we consider their association with Sgr tentative at best.

In conclusion, while there are likely stars from previously identified substructures in our sample of HiTS RRLs, it is unlikely that they have significantly biased our density profile fits, as each of the substructures would contribute only a small number of stars in our field of view.

\begin{table*}\scriptsize
\scriptsize
\caption{Parameters of number density profiles of the Halo from previous works, when power laws are considered. This Table only includes works with upper limits in distance  $>25$\,kpc.}
\label{tab:powerLaw}
\begin{center}
\begin{tabular}{|c c c c c c c c|}
\hline
\hline
	Model  &  Slope  &  $R_b$  &  Inner Slope  &  Outer slope & Range & Tracer Used & Paper\\
	 &    &   (kpc)    &    &    & (kpc) & &\\
\hline
Simple Power Law    &    &    &   &   &    &  &  \\  
  &  $-3.034\pm 0.08$ &  --  &  --  & -- & 1--80  & RRLs  &    \citealt{Wetterer1996}\\
  &  $\sim -2.8$ &  --  & -- & -- & 4--60  & RRLs  &   \citealt{Vivas2006}\\
    &  $-3.0$  &  --  &  --  &  --  &  5--40  & MSTO stars & \citealt{Bell2008}  \\
    &  $-2.5\pm 0.2$  &  --  &  --  &  --  &  10--90  & BHB stars & \citealt{DePropris2010}  \\
    &  $-2.7\pm 0.5$  &  --  &  --  &  --  &  1--40  & BHB and BS stars & \citealt{Deas11}  \\ 
   &  $-2.42\pm0.13$ &  --  & -- & --  & 5--30  & ($ab-$type) RRLs  &  \citealt{Sesar13a}\\
    &  $<-6$   &  --  &  --  &  --  &  50--100  & A-type stars & \citealt{Deas14}  \\ 
    &  $-3.8\pm 0.3$  &  --  &  --  &  --  &  50--100  & ($ab-$type) RRLs & \citealt{Cohen15}  \\
    &  $-3.4\pm 0.1$  &  --  &  --  &  --  &  10--80  & K giants & \citealt{Xue2015}  \\
    &  $-3.5\pm 0.2$  &  --  &  --  &  --  &  30--90  & Giant stars & \citealt{Slater2016}  \\
    &  $-2.96\pm 0.05$  &  --  &  -- & --  & 1--28  & RRLs & \citealt{Iorio2017}  \\
    &  \bm{$-4.17^{+0.18}_{-0.20}$}  &  --  &  --  &  --  &  \textbf{17--145}  & \textbf{RRLs} & \textbf{This work}  \\
\hline
Broken Power Law    &    &    &   &   &    &  &  \\ 
    &  --  &  $23$  &  $-2.4$  &  $-4.5$  &  5--100  & RRLs & \citealt{Wat09}  \\ 
    &  --  &  $27\pm 1$  &  $-2.3\pm 0.1$  &  $-4.6^{+0.2}_{-0.1}$  &  1--40  & BHB and BS stars & \citealt{Deas11}  \\ 
    &  --  &  $28$  &  $-2.6\pm 0.04$  &  $-3.8\pm 0.1$  &  12--40  & near MSTO stars & \citealt{Ses11}  \\ 
        &  -- &   $24$  &  $-2.8\pm 0.5$  &  $-5.4\pm 0.5$  &  5--60  & RRLs & \citealt{Zinn14}  \\ 
    &  --  &  $20$  &  $-2.5\pm 0.4$  &  $-4.9\pm 0.4$  &  10--60  & F-type stars & \citealt{PilaDiez2015}  \\ 
    &  --  &  $18\pm 1$  &  $-2.1\pm 0.3$  &  $-3.8\pm 0.1$  &  10--80  & K giants & \citealt{Xue2015}  \\ 
    &  --  &  $29.87^{+2.80}_{-3.55}$  &  $-3.61^{+0.15}_{-0.16}$  &  $-4.75^{+0.30}_{-0.28}$  &  10--70  & BHB stars & \citealt{Das2016}  \\ 
    &  --  &  \bm{$19.52^{+9.90}_{-3.27}$}  &  \bm{$-3.50^{+1.98}_{-1.46}$}  &  \bm{$-4.22^{+0.32}_{-0.26}$}  &  \textbf{17--145}  & \textbf{RRLs} & \textbf{This work}  \\
\hline
\end{tabular}
\end{center}
\end{table*}

\section{Discussion and Summary}
We present the detection of $173$ RRLs using observations from the 2014 campaign of the HiTS survey. The data cover $\sim 120$ square degrees of the sky and include from $20$ to $37$ epochs in the $g-$band. The photometric depth of the HiTS data enables us to build a catalog that includes a significant number of stars that do not appear in previous public surveys overlapping the same sky region (such as the CRTS). Most of the additional RRLs we contribute have $\left<g\right> > 20$, corresponding to $d_{\rm H} \gtrsim 60$\,kpc. 

The surveyed region overlaps with the position of currently known satellite galaxies of the Milky Way. We detect $65$ RRLs members of the Sextans dSph, $46$ of them being new discoveries. 
This number includes seven stars that were not classified as RRLs by other works, but were already considered as members of the dwarf. Two additional dwarf galaxies, Leo~IV and Leo~V, were found within our RRLs catalog.  
A more detailed analysis of the RRLs in Leo~V has been presented in a companion paper \citep{Medina2017}.

Regarding distant RRLs in our sample, we find $18$ candidates with $d_{\rm H}>90$\,kpc, $11$ of which are classified as $ab$-type, while seven are $c-$type. To understand the connection of this sub-sample with previously known/unknown systems, the period-amplitudes of the individual stars could play a major role. RRLs that are members of globular clusters are separated in the distincted Oosterhoff groups \citep[OoI and OoII,][]{Oosterhoff1939}. On the other hand, RRLs in ultra-faint dwarf galaxies are mostly OoII \citep{Clementini2014}, while the halo general population (within a radius of $\sim 80$ kpc) presents stars in both locus, but the majority of them follow the locus of the Oo I group. This has been shown multiple times in the literature \citep[e.g.][]{Zinn14,Catelan2015} and it is also clearly seen in Figure~\ref{fig:peramp}. The distant sample of RRLs however does not follow the main trend of nearby field RRLs, suggesting that the main contributor to the outer halo may come from UFD galaxies. 

We build number density radial profiles with our RRLs catalog (\S \ 5), then fit models of the form $\rho(R) = \rho_\odot(R/R_\odot)^n$ for a spherical halo and an ellipsoidal one with a flattening parameter $q=0.7$. The fits described in the previous section do not support the presence of a clear break in the profiles. However, due to the saturation limit of our photometry we only have $\sim1-2$ bins within the radius where a break is typically identified, so that our data may not be sensitive to the presence of a break. 
In fact, the data are well represented by a simple power law, with a slope of $-4.17^{+0.18}_{-0.20}$ for the spherical halo, and $n=-4.16^{+0.23}_{-0.25}$ for the elliptical case.
If a broken power law is considered, the profile (for a spherical halo) exhibits a break at $R_{b}=19.52^{+9.90}_{-3.27}$\,kpc, with inner and outer slopes of $n_{1}=-3.50^{+1.98}_{-1.46}$ and $n_{2}=-4.22^{+0.32}_{-0.26}$, respectively.
The position of the break and the values of the power-law indices are broadly consistent with previous studies.
We summarize values from the literature for these parameters in Table~\ref{tab:powerLaw}.
This agreement might represent a possible real feature of the Galactic halo, even when the sample used for those studies are located at different positions.
However, it is not clear that the data favor a certain model over the other. 
We tried a less likely double broken power law (a power law with two breaks) profile as well, but it was not considered for further analysis due to the lack of a strong physical meaning, big uncertainties and possible overfitting.

Our sample of HiTS RR Lyrae is an unprecedented probe of the outer Galactic halo; all but three of our stellar density bins are at distances beyond $30$\,kpc. As illustrated in Table~\ref{tab:powerLaw}, 
the power-law slope of the outer MW halo is typically found to be rather steep, whether it is measured by fitting a single power-law to the outer halo ($R_{\rm GC} \gtrsim 50$\,kpc), or a broken power-law with separate slopes for the inner/outer halo. The existence of a break in the halo density profile at $R_{\rm GC} \sim 20-30$\,kpc is typically argued to represent a transition between predominantly \textit{in situ} halo stars in the inner halo and an accretion-dominated outer halo \citep[e.g., ][]{BJ05,Abadi06}. \citet{Deas13} fit broken power-laws to the density profiles of synthetic, accretion-only models from \citet{BJ05}, and found a mean from the $11$ outer halo slopes of $\left<n\right> = -4.4$, with values ranging from $-2.5 > n > -6.5$, and median break radius of $26$\,kpc. Because all but two of our radial profile bins are beyond this typical break, these predictions provide a valid comparison to even our single power-law fits. \citet{Cooper2010} also predicts slightly steeper (on average) slopes (all with $n < -4.4$) for the outer halos of Aquarius model galaxies formed by the tidal disruption of dwarf satellites. The rough agreement of our power-law fits (both the SPL with $n = -4.17$, and the BPL with $n_{2} = -4.22, R_b = 19$\,kpc) with these predictions from halos made entirely of accreted satellites suggests that the outer MW halo may be made up completely of accreted stars.

We can perhaps learn even more about our Galaxy's accretion history based on the density profile. \citet{Pillepich14} fit the power-law stellar density profiles of $\sim5000$ MW analogs from the Illustris simulation suite between radii of the stellar half-mass radius ($r_{1/2}; \sim10$\,kpc for a $10^{12} M_\odot$ galaxy) out to the virial radius. These slopes range from $3.5 < n < 5.5$, with more massive halos exhibiting shallower stellar density profiles. More recently-formed halos, those that had a recent minor/major merger, or those that accreted a larger fraction of their stellar mass from satellites, typically have shallower slopes; thus, the fact that the MW's outer slope is often found to be steep suggests a relatively quiescent recent accretion history. Our measured slope of $n = -4.17$ is remarkably consistent with Illustris model predictions for MW-mass galaxies (see Figure 6 of \citealt{Pillepich14}).

With the addition of radial velocities for our outer halo RRLs, we can assess whether groups of stars which share similar radial velocities and 3D positions are part of physically associated, recently accreted tidal debris \citep[e.g., ][]{BakerWillman15,Vivas2016b}. If tidal debris structures in the outer halo can be linked to related structures at smaller Galactocentric radii, they provide a means of tracing the halo potential over the intervening distances \citep[e.g., ][]{Johnston12}. Furthermore, the $13$ RRLs we have discovered beyond $130$\,kpc in the Galaxy more than double the number of known stars at such distances in the MW. With spectroscopically-derived radial velocities, these stars will be vital tracers of the MW mass (e.g., \citealt{Watkins10,Sanderson16}), which is currently known only to roughly a factor of two \citep[e.g., ][]{Eadie2016, Ablimit2017}.

The data presented here cover only a small field of view in the Galactic halo, making interpretation within a global galaxy-formation context limited. Fully understanding our Galaxy's accretion history thus requires samples of tracers at $R_{\rm GC} > 100$\,kpc over many sky areas. Even with a handful of $\sim100$~deg$^2$ fields such as that presented in this work, we could examine the fraction of stars in substructures as a function of position, and the variation of density profiles with line of sight, while also combining data sets to average over local variations. The deep, $\sim20000$~deg$^2$, time-domain LSST survey \citep[e.g., ][]{LSST2009} will recover a nearly complete sample of RRLs out to $\gtrsim350$~kpc by the completion of the 10-year survey \citep[][see also Fig.~2 of \citealt{BakerWillman15}]{Ivezic08,Oluseyi12,VanderPlas15}. The HiTS results presented here are part of a larger ongoing LSST precursor observing program we are conducting, with which we will continue to map the outer limits of the Galactic halo with RR Lyrae variables, and develop tools for interpretation that lay the groundwork for exploiting the huge samples of RRLs in the LSST era.

\acknowledgments
We thank an anonymous referee for his/her comments and suggestions that helped improve this
manuscript.
G.M., F.F. and J.M. acknowledge support from the Ministry of Economy, Development, and Tourism's Millennium Science Initiative through grant IC120009, awarded to The Millennium Institute of Astrophysics (MAS), and from Conicyt through the Fondecyt Initiation into Research project No. 11130228. G.M. acknowledges CONICYT-PCHA/Mag\'isterNacional/2016-22162353. R.~R.~M.~acknowledges partial support from BASAL Project PFB-$06$ as well as FONDECYT project N$^{\circ}1170364$.
F.F. acknowledges support from BASAL Project PFB--03 and through the Programme of International Cooperation project DPI20140090. 
J.M. acknowledges the support from CONICYT Chile through CONICYT-PCHA/Doctorado-Nacional/2014-21140892, and Basal Project PFB-03, Centro de Modelamiento Matemáico (CMM), Universidad de Chile.
L.G. was supported in part by the US National Science Foundation under Grant AST-1311862. We acknowledge support from Conicyt through the infrastructure Quimal project No. 140003. Powered@NLHPC: this research was partially supported by the supercomputing infrastructure of the NLHPC (ECM-02). 
This project used data obtained with the Dark Energy Camera (DECam), which was constructed by the Dark Energy Survey (DES) collaboration. Funding for the DES Projects has been provided by the U.S. Department of Energy, the U.S. National Science Foundation, the Ministry of Science and Education of Spain, the Science and Technology Facilities Council of the United Kingdom, the Higher Education Funding Council for England, the National Center for Supercomputing Applications at the University of Illinois at Urbana-Champaign, the Kavli Institute of Cosmological Physics at the University of Chicago, Center for Cosmology and Astro-Particle Physics at the Ohio State University, the Mitchell Institute for Fundamental Physics and Astronomy at Texas A\&M University, Financiadora de Estudos e Projetos, Funda\c{c}\~ao Carlos Chagas Filho de Amparo, Financiadora de Estudos e Projetos, Funda\c{c}\~ao Carlos Chagas Filho de Amparo \`{a} Pesquisa do Estado do Rio de Janeiro, Conselho Nacional de Desenvolvimento Cient\'ifico e Tecnol\'ogico and the Minist\'erio da Ci\^{e}ncia, Tecnologia e Inova\c{c}\~ao, the Deutsche Forschungsgemeinschaft and the Collaborating Institutions in the Dark Energy Survey. The Collaborating Institutions are Argonne National Laboratory, the University of California at Santa Cruz, the University of Cambridge, Centro de Investigaciones En\'ergeticas, Medioambientales y Tecnol\'ogicas–Madrid, the University of Chicago, University College London, the DES-Brazil Consortium, the University of Edinburgh, the Eidgen\"{o}ssische Technische Hochschule (ETH) Z\"{u}rich, Fermi National Accelerator Laboratory, the University of Illinois at Urbana-Champaign, the Institut de Ci\`{e}ncies de l'Espai (IEEC/CSIC), the Institut de F\'isica d'Altes Energies, Lawrence Berkeley National Laboratory, the Ludwig-Maximilians Universit\"{a}t M\"{u}nchen and the associated Excellence Cluster Universe, the University of Michigan, the National Optical Astronomy Observatory, the University of Nottingham, the Ohio State University, the University of Pennsylvania, the University of Portsmouth, SLAC National Accelerator Laboratory, Stanford University, the University of Sussex, and Texas A\&M University.
R.R.M. acknowledges partial support from 
BASAL project PFB-$06$ as well as FONDECYT project N$^{\circ}1170364$. 

\newpage 

\appendix

\section{Tables and Lightcurves}
In this section, we include tables with the main properties of the sample of RRLs found by HiTS, as well as phased light curves for the stars that are not shown in Figure~\ref{fig:lcdistant}.
Table~\ref{tab:sextans} displays information about the RRLs in the Sextans dSph, while Table~\ref{tab:tab_general} shows the information of the rest of the sample.

\begin{table*}
\caption{RR Lyrae stars found in the Sextans dSph by HiTS. In the columun  Previous ID, stars with prefixes V and C are taken from \citet{Amigo2012}, with the former corresponding to detections made by \citet{Mateo1995}. The RRLs with prefixes LSQ are stars from LSQ \citep{Zinn14}, and the ones labeled as VV, VI and MV are from \citet{Lee03}. Stars with MV come from \citet{Mateo1995} as well. Note that only $4$ of our RRLs are classified as RRLs in \citet{Lee03}.}
\label{tab:sextans}
\begin{center}
\begin{tabular}{|c c c c c c c c c c|}
\hline
\hline
	ID  &  R.A.  &  DEC  &  $\left<g\right>$ & Period&  Amplitude  &  $d_{\rm H}$  & Type & N & Previous ID \\
	 &  (deg)  &  (deg)  & &  (days)  &   & (kpc) & & & \\
\hline
  HiTS100815-021625 & 152.06402 & -2.27369 & 20.3 & 0.6973 & 0.85 & 84  & ab & 30 & - \\
  HiTS100936-015356 & 152.40198 & -1.89876 & 20.4 & 0.346 & 0.65 & 84  & c & 38 & - \\
  HiTS100958-015954 & 152.49002 & -1.99832 & 20.4 & 0.6497 & 0.97 & 84  & ab & 36 & - \\
  HiTS101010-021303 & 152.54151 & -2.21761 & 20.4 & 0.6775 & 0.65 & 87  & ab & 37 & - \\
  HiTS101011-012028 & 152.54492 & -1.34107 & 20.4 & 0.675 & 0.67 & 86  & ab & 38 & - \\
  HiTS101035-014902 & 152.64737 & -1.81732 & 20.4 & 0.728 & 0.56 & 87  & ab & 38 & - \\
  HiTS101059-014146 & 152.74635 & -1.69613 & 20.4 & 0.60505 & 0.87 & 85  & ab & 37 & - \\
  HiTS101104-021031 & 152.76652 & -2.17535 & 20.4 & 0.5939 & 0.94 & 84  & ab & 37 & - \\
  HiTS101110-022812 & 152.79187 & -2.46998 & 20.3 & 0.3059 & 0.46 & 79  & c & 35 & - \\
  HiTS101118-014123 & 152.82598 & -1.68969 & 20.4 & 0.6088 & 0.69 & 85  & ab & 33 & - \\
  HiTS101123-013813 & 152.84444 & -1.63696 & 20.4 & 0.6031 & 0.8 & 83  & ab & 38 & - \\
  HiTS101127-021833 & 152.86234 & -2.30928 & 20.3 & 0.7153 & 0.69 & 85  & ab & 37 & - \\
  HiTS101137-013159 & 152.90356 & -1.53294 & 20.4 & 0.3941 & 0.57 & 90  & c & 37 & - \\
  HiTS101146-014207 & 152.94133 & -1.70199 & 20.2 & 0.3163 & 0.69 & 76  & c & 38 & VV23\\
  HiTS101147-012921 & 152.94423 & -1.48912 & 20.5 & 0.3012 & 0.64 & 83  & c & 37 & VI02\\
  HiTS101149-023558 & 152.95618 & -2.59951 & 20.2 & 0.6395 & 1.04 & 79  & ab & 20 & - \\
  HiTS101159-020215 & 152.99415 & -2.03763 & 20.0 & 0.4601 & 0.53 & 80  & c & 37 & - \\
  HiTS101159-014352 & 152.99676 & -1.73099 & 20.4 & 0.3472 & 0.75 & 84  & c & 38 & C5\\
  HiTS101200-020233 & 152.99877 & -2.04241 & 20.5 & 0.5999 & 1.0 & 87  & ab & 37 & - \\
  HiTS101214-014048 & 153.05782 & -1.68014 & 20.5 & 0.6093 & 0.63 & 87  & ab & 38 & C13,VI47\\
  HiTS101215-014921 & 153.06118 & -1.82237 & 20.4 & 0.66581 & 0.71 & 86  & ab & 38 & C14,VV26\\
  HiTS101215-015415 & 153.06268 & -1.90417 & 20.3 & 0.4261 & 0.51 & 90  & c & 38 & C15\\
  HiTS101215-013710 & 153.06378 & -1.61944 & 20.6 & 0.629 & 0.82 & 90  & ab & 38 & C16,VI46\\
  HiTS101216-020415 & 153.06648 & -2.07081 & 20.5 & 0.6532 & 0.85 & 88  & ab & 37 & - \\
  HiTS101218-020021 & 153.07314 & -2.00582 & 20.3 & 0.73271 & 0.75 & 83  & ab & 37 & LSQ27\\
  HiTS101219-014025 & 153.07974 & -1.67363 & 20.4 & 0.5937 & 0.93 & 83  & ab & 38 & C21,VI44\\
  HiTS101219-015445 & 153.08119 & -1.91262 & 20.3 & 0.3776 & 0.57 & 85  & c & 33 & - \\
  HiTS101220-014915 & 153.08398 & -1.82097 & 20.3 & 0.5817 & 1.03 & 78  & ab & 38 & C23\\
  HiTS101221-013733 & 153.08562 & -1.62578 & 20.4 & 0.7032 & 0.86 & 88  & ab & 38 & C24,VI45\\
  HiTS101223-015225 & 153.09601 & -1.87375 & 20.4 & 0.5828 & 1.07 & 81  & ab & 37 & C30\\
  HiTS101225-014319 & 153.10533 & -1.72193 & 20.5 & 0.5726 & 1.05 & 84  & ab & 37 & C33,VI49\\
  HiTS101249-020327 & 153.20288 & -2.05756 & 20.5 & 0.6361 & 0.74 & 87  & ab & 37 & - \\
  HiTS101309-020442 & 153.28603 & -2.07826 & 20.2 & 0.6606 & 0.63 & 79  & ab & 20 & LSQ33\\
  HiTS101312-021412 & 153.30010 & -2.23653 & 20.4 & 0.627 & 0.78 & 84  & ab & 21 & - \\
  HiTS101316-020029 & 153.31510 & -2.00796 & 20.5 & 0.6105 & 0.79 & 87  & ab & 21 & - \\
  HiTS101333-003607 & 153.38637 & -0.60196 & 20.2 & 0.6626 & 1.03 & 81  & ab & 21 & - \\
  HiTS101336-020026 & 153.40018 & -2.00731 & 20.3 & 0.6052 & 1.13 & 80  & ab & 21 & - \\
  HiTS101337-014947 & 153.40393 & -1.82963 & 20.5 & 0.6285 & 0.78 & 88  & ab & 21 & C102,VI78\\
  HiTS101339-014310 & 153.41355 & -1.71934 & 20.4 & 0.3464 & 0.65 & 85  & c & 20 & V22,VV35(RRL)\\
  HiTS101344-014845 & 153.43204 & -1.81237 & 20.4 & 0.5887 & 0.97 & 83  & ab & 21 & V25,VI75(RRL)\\
  HiTS101345-014305 & 153.43776 & -1.71792 & 20.5 & 0.6215 & 0.78 & 87  & ab & 20 & V26,VI80(RRL)\\
  HiTS101349-015042 & 153.45497 & -1.84497 & 20.3 & 0.3962 & 0.53 & 87  & c & 21 & VI77\\
  HiTS101356-004811 & 153.48130 & -0.80309 & 20.4 & 0.3616 & 0.61 & 88  & c & 21 & - \\
  HiTS101401-014046 & 153.50327 & -1.67939 & 20.5 & 0.48777 & 1.13 & 82  & ab & 20 & V42,MV09(RRL)\\
  HiTS101403-013845 & 153.51083 & -1.64578 & 20.4 & 0.5417 & 1.08 & 82  & ab & 20 & V43\\
  HiTS101413-004502 & 153.55313 & -0.75058 & 20.2 & 0.3861 & 0.62 & 84  & c & 21 & LSQ39\\
  HiTS101413-014709 & 153.55455 & -1.78591 & 20.4 & 0.6149 & 0.94 & 84  & ab & 21 & VI81\\
  HiTS101413-022107 & 153.55496 & -2.35185 & 20.2 & 0.3032 & 0.49 & 76  & c & 21 & - \\
  HiTS101414-013033 & 153.55636 & -1.50920 & 20.4 & 0.6342 & 0.63 & 85  & ab & 20 & VV21\\
  HiTS101415-015358 & 153.56291 & -1.89934 & 20.3 & 0.6141 & 0.68 & 81  & ab & 21 & - \\
  HiTS101417-015904 & 153.57276 & -1.98455 & 20.4 & 0.5919 & 0.96 & 82  & ab & 21 & - \\
  HiTS101423-012721 & 153.59391 & -1.45573 & 20.3 & 0.3442 & 0.52 & 82  & c & 21 & - \\
  HiTS101426-014200 & 153.60822 & -1.69990 & 20.4 & 0.5424 & 0.97 & 82  & ab & 20 & VI83\\
  HiTS101426-014048 & 153.60920 & -1.68002 & 20.5 & 0.6156 & 0.77 & 87  & ab & 20 & VI84\\
  HiTS101428-014334 & 153.61674 & -1.72623 & 20.4 & 0.54928 & 1.08 & 79  & ab & 20 & - \\
  HiTS101429-004613 & 153.62042 & -0.77035 & 20.5 & 0.6324 & 0.8 & 87  & ab & 21 & - \\
  HiTS101446-012748 & 153.69002 & -1.46334 & 20.5 & 0.6074 & 0.94 & 86  & ab & 21 & - \\
  HiTS101447-012724 & 153.69536 & -1.45670 & 20.4 & 0.3682 & 0.52 & 88  & c & 21 & - \\
  HiTS101500-013822 & 153.75073 & -1.63955 & 20.3 & 0.7754 & 0.49 & 85  & ab & 21 & - \\
  HiTS101511-011825 & 153.79493 & -1.30695 & 20.3 & 0.7262 & 0.62 & 85  & ab & 21 & - \\
  HiTS101514-013400 & 153.80651 & -1.56672 & 20.5 & 0.60825 & 0.86 & 87  & ab & 21 & - \\
  HiTS101551-013732 & 153.96215 & -1.62542 & 20.2 & 0.6771 & 0.9 & 81  & ab & 21 & - \\
  HiTS101605-012847 & 154.01876 & -1.47980 & 20.4 & 0.6136 & 0.52 & 83  & ab & 21 & - \\
  HiTS101626-013756 & 154.10991 & -1.63218 & 20.3 & 0.6585 & 0.94 & 81  & ab & 21 & - \\
  HiTS101638-011903 & 154.15994 & -1.31752 & 20.4 & 0.407 & 0.46 & 90  & c & 21 & - \\
\hline
\end{tabular} 
\end{center}
\end{table*}

\newpage 

\begin{longtable}{|c c c c c c c c c|}
\caption{Full list of the RRLs presented in this work, excluding the candidates in the Sextans dSph galaxy.}
\label{tab:tab_general}
\endfirsthead
\endfoot
\caption*{Continued.}\\\toprule
	ID  &  R.A.  &  DEC  &   $\left<g\right>$ &  Period  & Amplitude & $d_{\rm H}$  & Type & N \\
	 &  (deg)  &  (deg)  &  & (days)  &   &  (kpc) &  & \\
\hline
\endhead
\hline
\endfoot 
\hline
\hline
	ID  &  R.A.  &  DEC  &   $\left<g\right>$ &  Period  & Amplitude & $d_{\rm H}$  & Type & N \\
	 &  (deg)  &  (deg)  &  & (days)  &   &  (kpc) &  & \\
\hline
  HiTS095845+023118 & 149.68616 & 2.52172 & 17.9 & 0.6453 & 0.36 & 28  & ab & 18\\
  HiTS100129+001100 & 150.37090 & 0.18338 & 16.4 & 0.6098 & 1.30 & 13  & ab & 21\\
  HiTS100207+012232 & 150.52751 & 1.37554 & 16.5 & 0.7208 & 0.91 & 15  & ab & 16\\
  HiTS100236+030023 & 150.64878 & 3.00649 & 16.1 & 0.6079 & 1.02 & 12  & ab & 20\\
  HiTS100838-020312 & 152.15703 & -2.05345 & 17.4 & 0.5799 & 1.28 & 20  & ab & 38\\
  HiTS100843-041555 & 152.17745 & -4.26539 & 17.9 & 0.6215 & 0.48 & 27  & ab & 21\\
  HiTS100911-041232 & 152.29475 & -4.20889 & 17.8 & 0.3754 & 0.65 & 26  & c & 21\\
  HiTS100942+012903 & 152.42563 & 1.48407 & 19.3 & 0.4875 & 0.81 & 47  & ab & 20\\
  HiTS100956+013212 & 152.48158 & 1.53665 & 20.4 & 0.6220 & 0.68 & 84  & ab & 20\\
  HiTS101014-020114 & 152.55937 & -2.02065 & 16.8 & 0.5574 & 1.08 & 15  & ab & 35\\
  HiTS101057-033322 & 152.73691 & -3.55624 & 19.8 & 0.6363 & 1.17 & 63  & ab & 21\\
  HiTS101243+022118 & 153.17763 & 2.35503 & 20.6 & 0.5346 & 1.04 & 87  & ab & 20\\
  HiTS101321-035610 & 153.33614 & -3.93599 & 18.6 & 0.5871 & 0.97 & 36  & ab & 21\\
  HiTS101336-013752 & 153.40203 & -1.63120 & 18.8 & 0.6960 & 0.88 & 42  & ab & 20\\
  HiTS101353-035021 & 153.46884 & -3.83912 & 17.3 & 0.4162 & 0.23 & 22  & c & 21\\
  HiTS101452-002635 & 153.71870 & -0.44305 & 18.1 & 0.5803 & 0.96 & 28  & ab & 21\\
  HiTS101529-012126 & 153.86970 & -1.35711 & 20.1 & 0.4132 & 0.71 & 63  & ab & 21\\
  HiTS101626-041658 & 154.10903 & -4.28284 & 17.6 & 0.7404 & 0.58 & 24  & ab & 21\\
  HiTS101750-020127 & 154.45848 & -2.02430 & 16.5 & 0.3307 & 0.46 & 14  & c & 21\\
  HiTS101936-012754 & 154.89982 & -1.46512 & 17.1 & 0.6329 & 0.51 & 18  & ab & 21\\
  HiTS101953-050553 & 154.97160 & -5.09813 & 19.3 & 0.6865 & 0.82 & 50  & ab & 20\\
  HiTS101957-020302 & 154.98816 & -2.05048 & 19.5 & 0.4292 & 0.95 & 48  & ab & 21\\
  HiTS102014-042354 & 155.05789 & -4.39843 & 22.5 & 0.3841 & 0.37 & 233  & c & 19\\
  HiTS102106-064439 & 155.27585 & -6.74404 & 18.9 & 0.5985 & 0.94 & 41  & ab & 17\\
  HiTS102306-080748 & 155.77552 & -8.12994 & 17.4 & 0.4663 & 1.40 & 18  & ab & 21\\
  HiTS102313-082435 & 155.80301 & -8.40973 & 17.2 & 0.4632 & 1.30 & 17  & ab & 20\\
  HiTS102344-064233 & 155.93215 & -6.70930 & 18.2 & 0.6582 & 0.45 & 31  & ab & 17\\
  HiTS102414-095518 & 156.05905 & -9.92180 & 21.7 & 0.7641 & 0.54 & 161  & ab & 21\\
  HiTS102610-083620 & 156.54047 & -8.60557 & 16.2 & 0.3228 & 0.53 & 12  & c & 22\\
  HiTS103015-050203 & 157.56265 & -5.03403 & 17.1 & 0.3032 & 0.62 & 18  & c & 22\\
  HiTS103601-015451 & 159.00456 & -1.91422 & 21.7 & 0.4045 & 0.29 & 161  & c & 21\\
  HiTS103626-014703 & 159.10785 & -1.78424 & 16.1 & 0.5872 & 1.08 & 11  & ab & 19\\
  HiTS103758-043930 & 159.49133 & -4.65828 & 17.9 & 0.5836 & 0.89 & 25  & ab & 21\\
  HiTS103931-034037 & 159.87731 & -3.67708 & 17.1 & 0.5480 & 0.70 & 17  & ab & 21\\
  HiTS103943-021726 & 159.93119 & -2.29061 & 20.9 & 0.6956 & 0.44 & 108  & ab & 21\\
  HiTS104009-063304 & 160.03895 & -6.55105 & 20.8 & 0.6376 & 0.69 & 100  & ab & 21\\
  HiTS104054-042827 & 160.22661 & -4.47424 & 21.9 & 0.4650 & 0.34 & 183  & c & 20\\
  HiTS104402-040641 & 161.00852 & -4.11148 & 17.4 & 0.5705 & 1.04 & 20  & ab & 21\\
  HiTS104407-035317 & 161.03109 & -3.88797 & 18.0 & 0.6190 & 0.63 & 28  & ab & 21\\
  HiTS104423+011722 & 161.09545 & 1.28950 & 16.4 & 0.3620 & 0.51 & 13  & c & 21\\
  HiTS104427+024346 & 161.11385 & 2.72931 & 17.6 & 0.3198 & 0.40 & 23  & c & 18\\
  HiTS104738+020627 & 161.90718 & 2.10746 & 22.8 & 0.3844 & 0.47 & 262  & c & 19\\
  HiTS104924-023635 & 162.34954 & -2.60969 & 17.5 & 0.5263 & 1.28 & 21  & ab & 20\\
  HiTS105142+001611 & 162.92606 & 0.26978 & 16.5 & 0.6488 & 1.07 & 14  & ab & 20\\
  HiTS105152+021339 & 162.96779 & 2.22743 & 16.1 & 0.5548 & 1.12 & 11  & ab & 20\\
  HiTS105209-043942 & 163.03718 & -4.66174 & 21.5 & 0.6036 & 0.46 & 136  & ab & 20\\
  HiTS105213-043434 & 163.05435 & -4.57609 & 17.3 & 0.3632 & 0.17 & 21  & c & 20\\
  HiTS105226+022759 & 163.10707 & 2.46631 & 16.8 & 0.5584 & 1.07 & 16  & ab & 21\\
  HiTS105311+015046 & 163.29669 & 1.84613 & 17.2 & 0.2893 & 0.28 & 18  & c & 20\\
  HiTS105502-025906 & 163.75656 & -2.98510 & 16.9 & 0.5797 & 0.94 & 16  & ab & 20\\
  HiTS105537+022053 & 163.90258 & 2.34818 & 17.0 & 0.5476 & 0.90 & 17  & ab & 21\\
  HiTS105545-013021 & 163.93648 & -1.50583 & 16.5 & 0.4599 & 1.34 & 12  & ab & 20\\
  HiTS105545-013440 & 163.93662 & -1.57783 & 20.4 & 0.5219 & 1.01 & 76  & ab & 20\\
  HiTS105549+023408 & 163.95249 & 2.56879 & 17.0 & 0.4496 & 1.44 & 15  & ab & 21\\
  HiTS105613-015658 & 164.05547 & -1.94939 & 18.1 & 0.7732 & 0.41 & 30  & ab & 20\\
  HiTS105634+013335 & 164.14314 & 1.55986 & 18.1 & 0.3041 & 0.53 & 29  & c & 20\\
  HiTS105726+024352 & 164.35679 & 2.73115 & 19.4 & 0.5843 & 1.02 & 53  & ab & 21\\
  HiTS105738+001316 & 164.41037 & 0.22123 & 18.2 & 0.6421 & 0.53 & 30  & ab & 21\\
  HiTS105754-002603 & 164.47577 & -0.43403 & 20.5 & 0.3609 & 0.62 & 93  & c & 20\\
  HiTS105843+023125 & 164.67737 & 2.52354 & 18.0 & 0.6988 & 0.71 & 29  & ab & 21\\
  HiTS105933+020617 & 164.88765 & 2.10477 & 16.9 & 0.4653 & 1.31 & 15  & ab & 19\\
  HiTS110117-035105 & 165.32027 & -3.85142 & 19.9 & 0.4277 & 0.53 & 57  & ab & 18\\
  HiTS110200+005324 & 165.49980 & 0.89006 & 17.6 & 0.5336 & 0.57 & 23  & ab & 21\\
  HiTS110222-001624 & 165.59251 & -0.27337 & 22.1 & 0.6118 & 0.51 & 180  & ab & 19\\
  HiTS110243+014821 & 165.68036 & 1.80572 & 17.2 & 0.7910 & 0.55 & 21  & ab & 20\\
  HiTS110254-003415 & 165.72515 & -0.57075 & 17.4 & 0.3383 & 0.48 & 22  & c & 21\\
  HiTS110456+015745 & 166.23204 & 1.96261 & 17.2 & 0.6505 & 0.93 & 20  & ab & 20\\
  HiTS110510-022710 & 166.28982 & -2.45282 & 22.4 & 0.7459 & 0.61 & 219  & ab & 19\\
  HiTS110616+015120 & 166.56478 & 1.85545 & 16.4 & 0.3509 & 0.48 & 14  & c & 20\\
  HiTS110739+012813 & 166.91383 & 1.47037 & 20.4 & 0.4359 & 0.24 & 93  & c & 20\\
  HiTS110829+013851 & 167.11962 & 1.64753 & 16.5 & 0.5937 & 1.05 & 14  & ab & 20\\
  HiTS111033-032936 & 167.63857 & -3.49333 & 15.6 & 0.3439 & 0.49 & 9  & c & 22\\
  HiTS111106-041718 & 167.77512 & -4.28834 & 21.2 & 0.3034 & 0.62 & 113  & c & 22\\
  HiTS111246-043641 & 168.19260 & -4.61130 & 16.5 & 0.5258 & 1.13 & 13  & ab & 21\\
  HiTS111311+025109 & 168.29469 & 2.85259 & 16.5 & 0.7284 & 0.67 & 15  & ab & 19\\
  HiTS111342-021922 & 168.42638 & -2.32288 & 17.3 & 0.3103 & 0.44 & 20  & c & 22\\
  HiTS111420+024911 & 168.58268 & 2.81958 & 17.6 & 0.5540 & 1.10 & 22  & ab & 20\\
  HiTS111522-025143 & 168.84248 & -2.86206 & 18.0 & 0.6350 & 0.58 & 27  & ab & 21\\
  HiTS111722+022620 & 169.33980 & 2.43886 & 17.2 & 0.6322 & 0.99 & 19  & ab & 20\\
  HiTS111934-041741 & 169.89346 & -4.29472 & 15.8 & 0.6413 & 0.37 & 10  & ab & 22\\
  HiTS112000-003831 & 169.99881 & -0.64205 & 16.6 & 0.4494 & 0.99 & 13  & ab & 20\\
  HiTS112039-040702 & 170.16125 & -4.11709 & 16.2 & 0.3906 & 0.55 & 13  & c & 22\\
  HiTS112058-020326 & 170.23962 & -2.05724 & 16.4 & 0.3516 & 0.49 & 14  & c & 22\\
  HiTS112445+013726 & 171.18809 & 1.62395 & 18.1 & 0.5997 & 0.81 & 28  & ab & 20\\
  HiTS112524-024348 & 171.35051 & -2.72999 & 20.4 & 0.6373 & 0.43 & 83  & ab & 22\\
  HiTS112549-041215 & 171.45363 & -4.20410 & 15.5 & 0.6733 & 0.84 & 9  & ab & 20\\
  HiTS112807-034733 & 172.03057 & -3.79243 & 18.2 & 0.6359 & 1.14 & 29  & ab & 20\\
  HiTS112820-000636 & 172.08521 & -0.11014 & 19.3 & 0.6840 & 0.74 & 54  & ab & 20\\
  HiTS112838-000112 & 172.15740 & -0.02008 & 18.9 & 0.5699 & 1.10 & 40  & ab & 20\\
  HiTS112859-034415 & 172.24634 & -3.73761 & 20.4 & 0.3934 & 0.52 & 84  & c & 20\\
  HiTS113006+001108 & 172.52405 & 0.18565 & 17.9 & 0.6584 & 0.28 & 28  & ab & 20\\
  HiTS113057+021331 & 172.73946 & 2.22514 & 21.8 & 0.6453 & 0.72 & 172  & ab & 20\\
  HiTS113105+021319 & 172.76936 & 2.22200 & 21.8 & 0.6574 & 1.40 & 172  & ab & 20\\
  HiTS113107+023025 & 172.77770 & 2.50700 & 18.8 & 0.6626 & 0.90 & 41  & ab & 21\\
  HiTS113107+021302 & 172.77796 & 2.21734 & 21.7 & 0.6442 & 1.02 & 163  & ab & 21\\
  HiTS113217-035542 & 173.07262 & -3.92844 & 17.9 & 0.7370 & 0.53 & 27  & ab & 20\\
  HiTS113256-003329 & 173.23270 & -0.55818 & 21.5 & 0.7146 & 0.78 & 156  & ab & 19\\
  HiTS113259-003404 & 173.24674 & -0.56770 & 21.5 & 0.6268 & 1.22 & 147  & ab & 18\\
  HiTS113314+022239 & 173.30754 & 2.37736 & 17.6 & 0.6029 & 0.69 & 23  & ab & 20\\
  HiTS113336-011012 & 173.39867 & -1.17007 & 16.9 & 0.5125 & 1.21 & 16  & ab & 20\\
  HiTS113353+010816 & 173.46942 & 1.13774 & 17.3 & 0.8128 & 0.83 & 23  & ab & 21\\
  HiTS113400+024753 & 173.49853 & 2.79812 & 18.7 & 0.6703 & 0.86 & 39  & ab & 21\\
  HiTS113600+021833 & 173.99975 & 2.30909 & 17.8 & 0.6626 & 0.61 & 27  & ab & 18\\
  HiTS113609+025834 & 174.03651 & 2.97606 & 17.7 & 0.5505 & 1.01 & 24  & ab & 21\\
  HiTS113634-012016 & 174.14059 & -1.33785 & 18.7 & 0.6242 & 0.58 & 39  & ab & 20\\
  HiTS113640-012516 & 174.16589 & -1.42113 & 16.8 & 0.6507 & 0.60 & 17  & ab & 20\\
  HiTS113653-024629 & 174.22135 & -2.77481 & 16.7 & 0.2668 & 0.40 & 14  & c & 20\\
  HiTS113748-021817 & 174.45040 & -2.30472 & 16.8 & 0.5724 & 1.11 & 16  & ab & 20\\
\hline
\end{longtable}

\newpage

\begin{figure*}
\includegraphics[width=17cm]{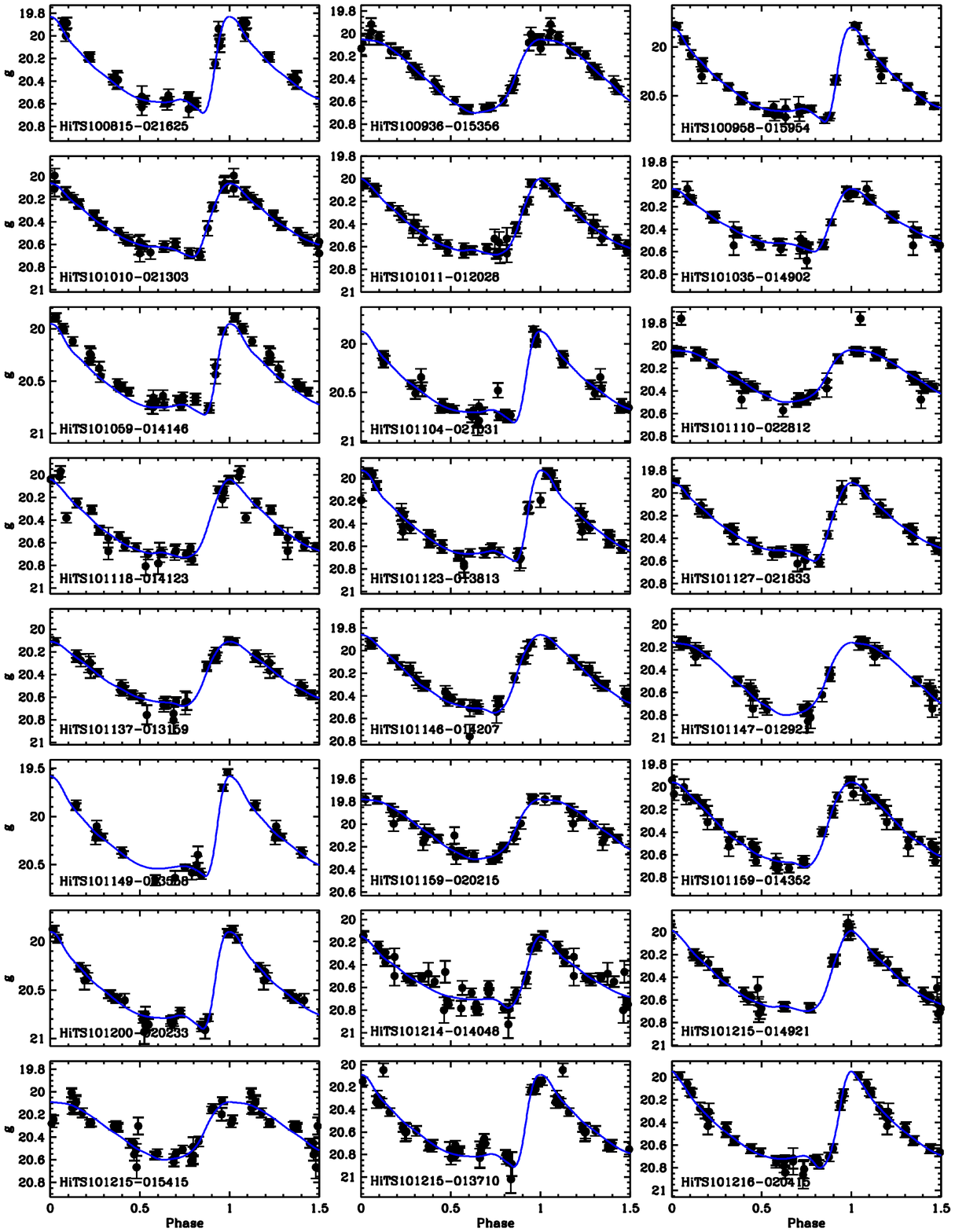}
\caption{Phased light curves of RR Lyrae stars in the Sextans dwarf spheroidal galaxy (1/3).}
\label{fig:lcSextans1}
\end{figure*}

\begin{figure*}
\includegraphics[width=17cm]{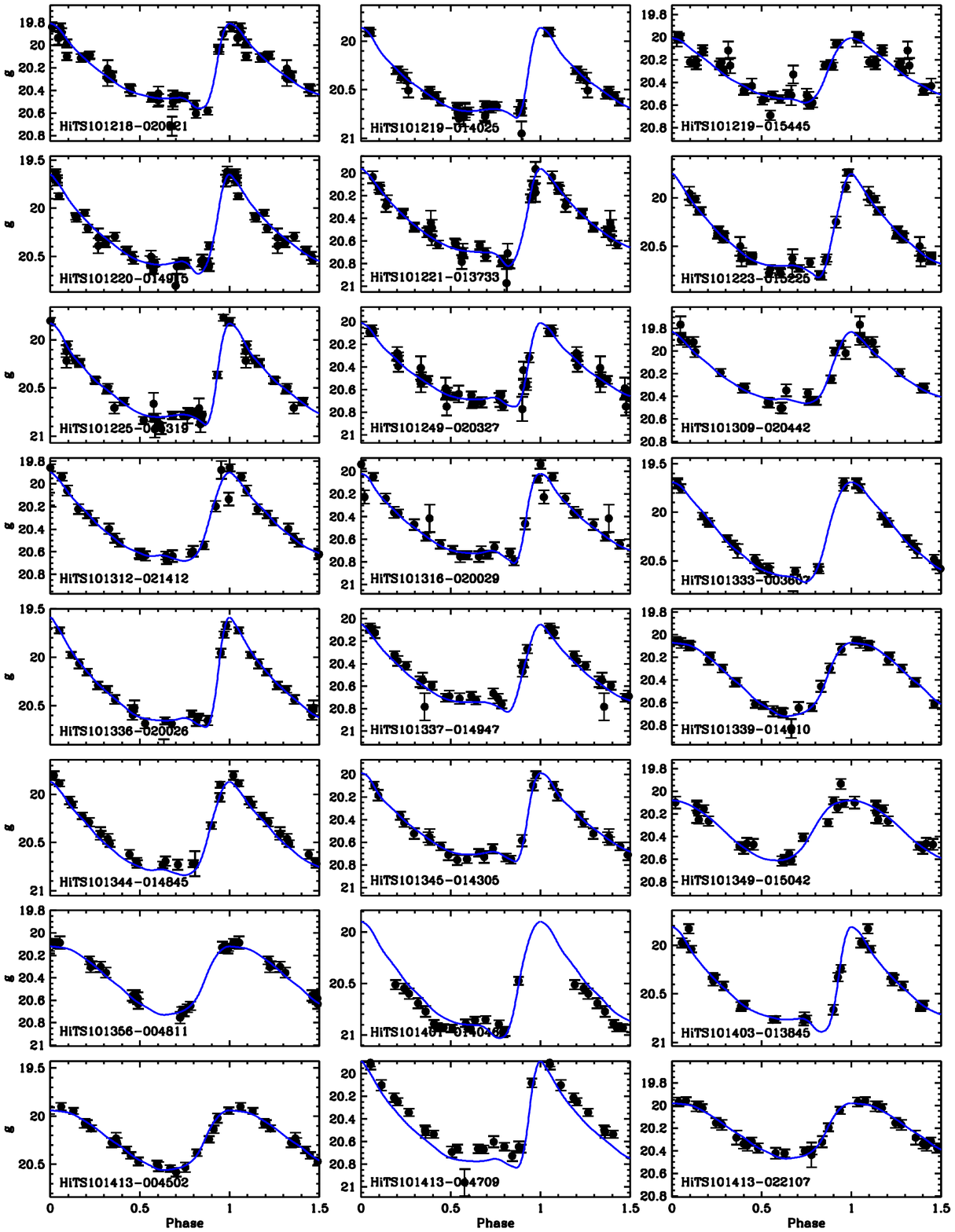}
\caption{Phased light curves of RR Lyrae stars in the Sextans dwarf spheroidal galaxy (2/3).}
\label{fig:lcSextans2}
\end{figure*}

\begin{figure*}
\includegraphics[width=17cm]{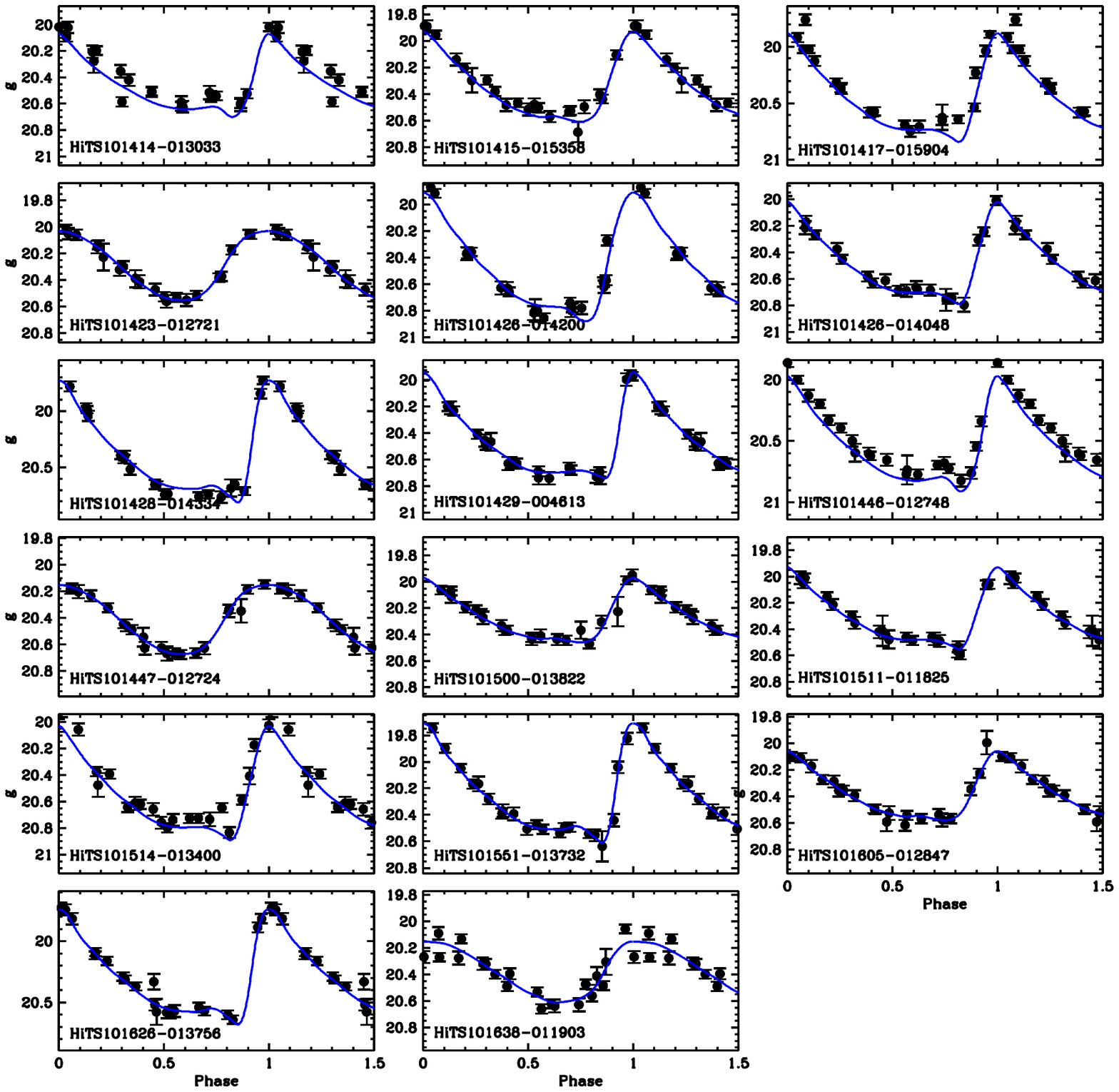}
\caption{Phased light curves of RR Lyrae stars in the Sextans dwarf spheroidal galaxy (3/3).}
\label{fig:lcSextans3}
\end{figure*}

\begin{figure*}
\includegraphics[width=17cm]{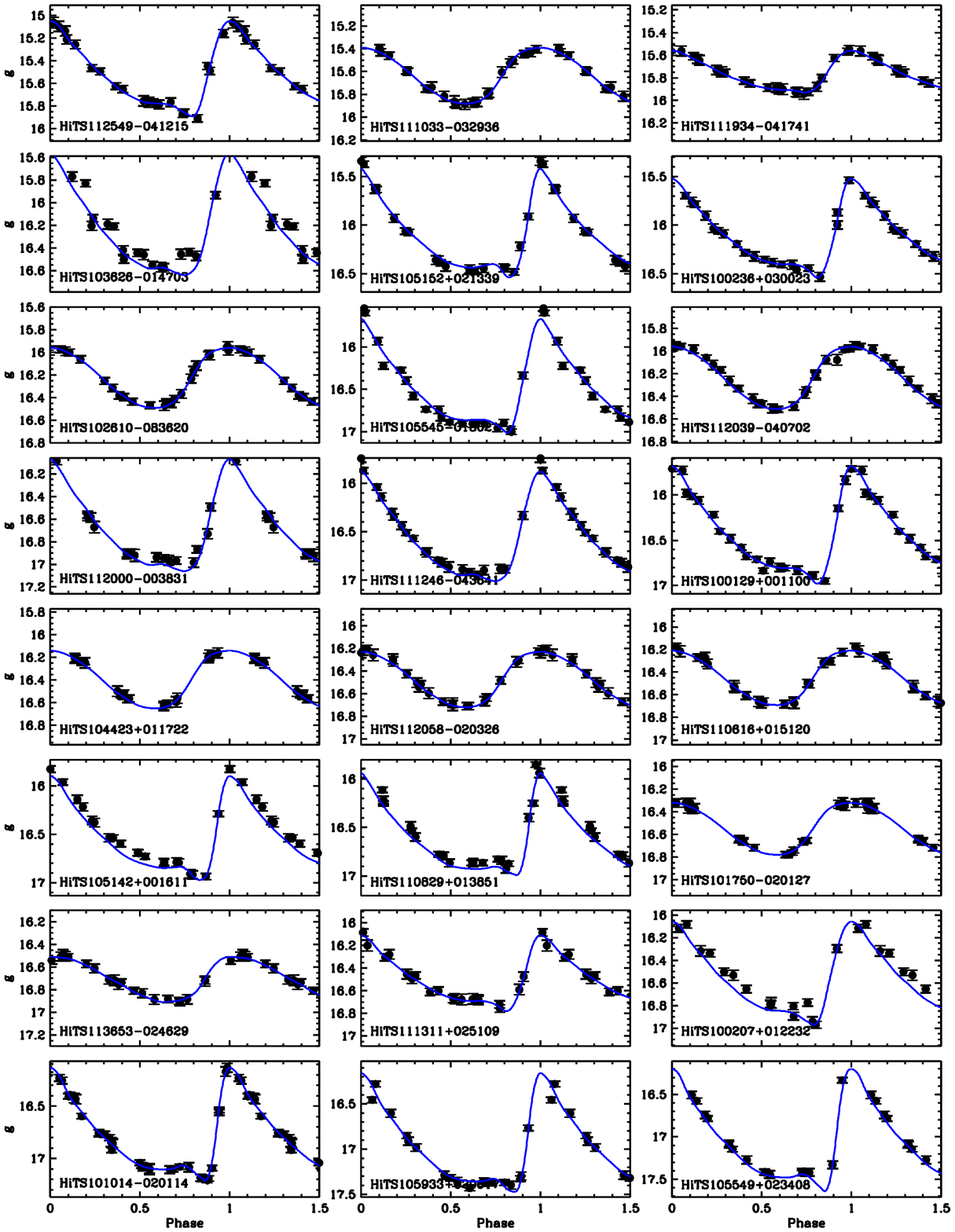}
\caption{Phased light curves of nearby ($<90$\,kpc) RR Lyrae stars in the field (1/4).}
\label{fig:lcField1}
\end{figure*}

\begin{figure*}
\includegraphics[width=17cm]{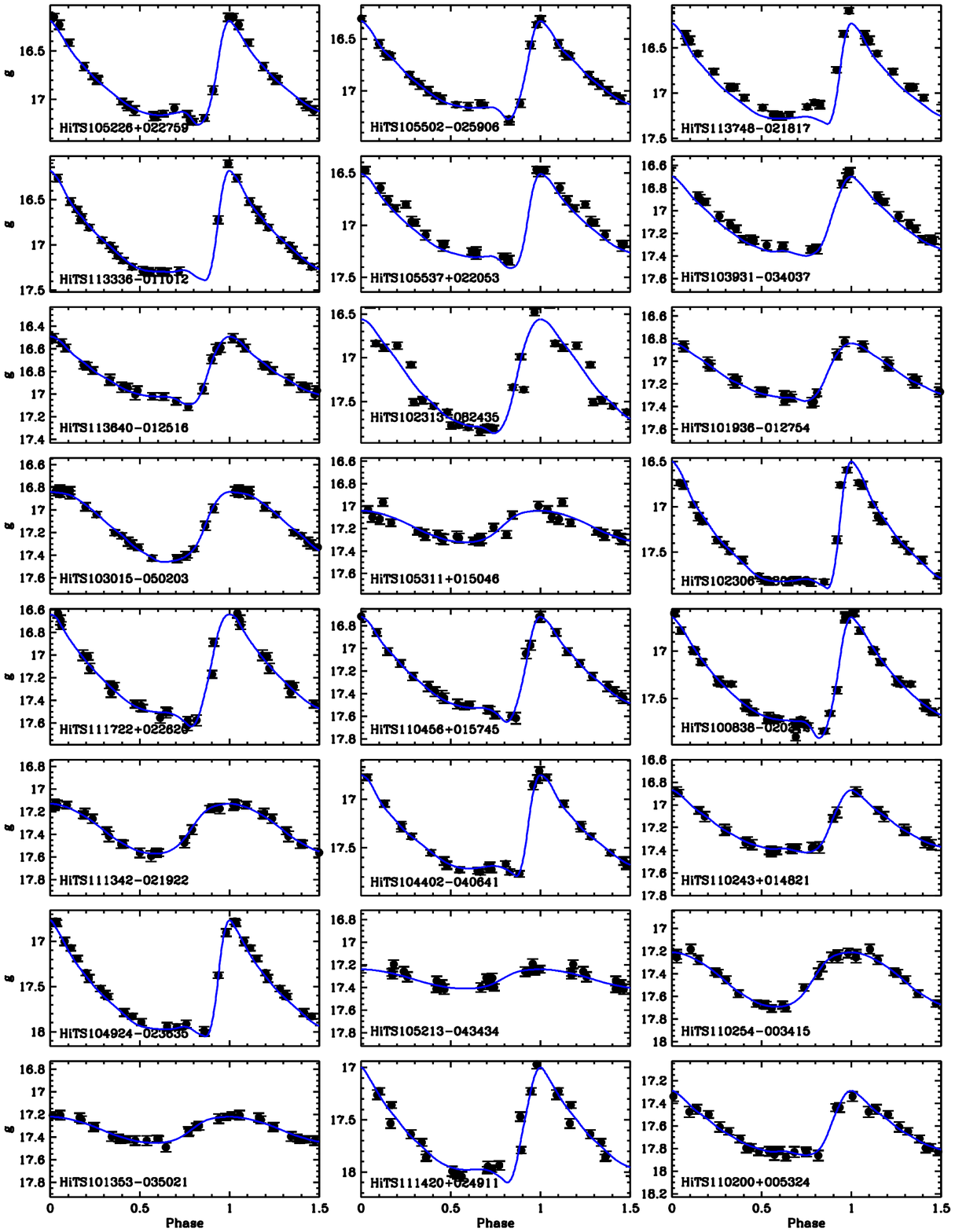}
\caption{Phased light curves of nearby ($<90$\,kpc) RR Lyrae stars in the field (2/4).}
\label{fig:lcField2}
\end{figure*}

\begin{figure*}
\includegraphics[width=17cm]{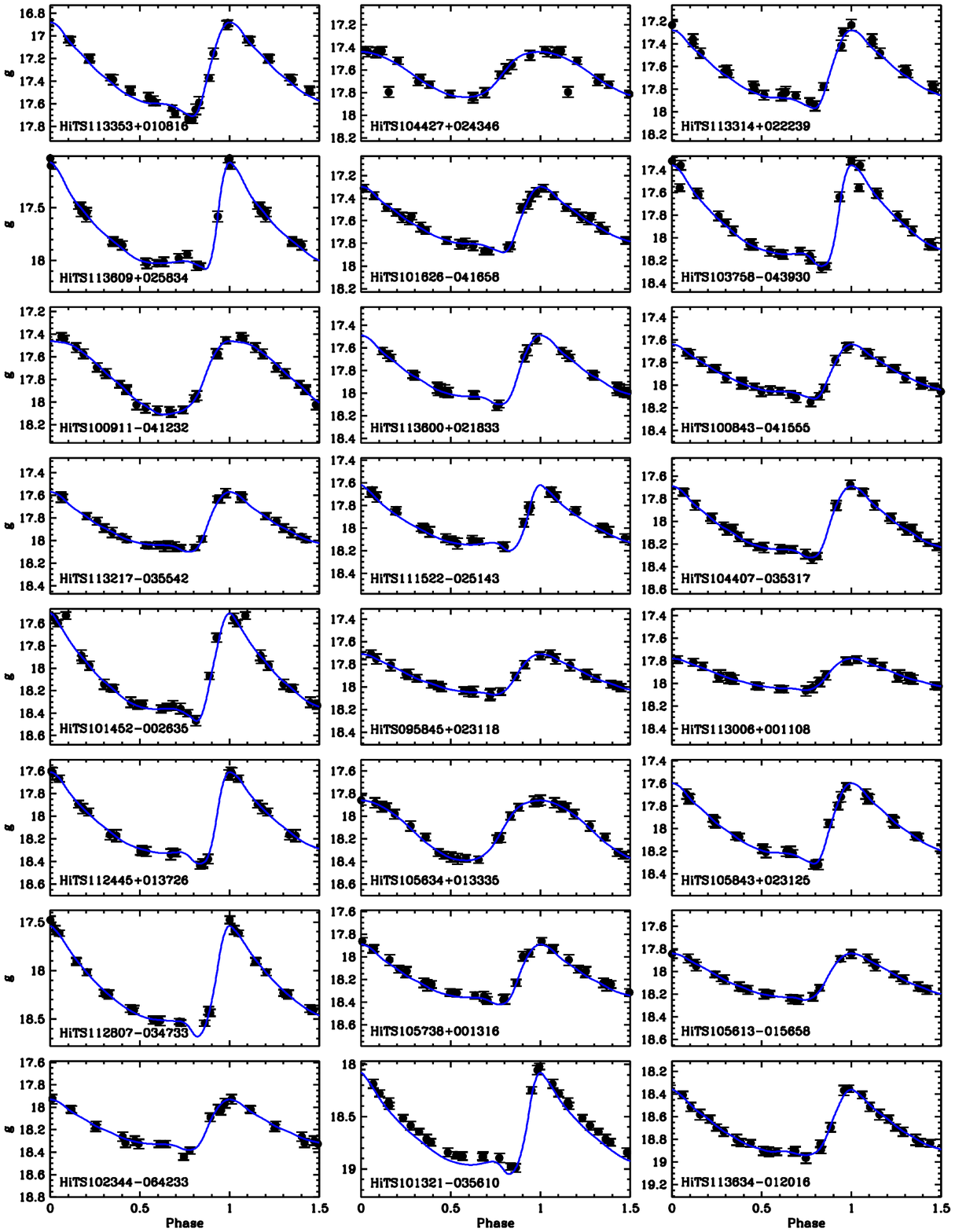}
\caption{Phased light curves of nearby ($<90$\,kpc) RR Lyrae stars in the field (3/4).}
\label{fig:lcField3}
\end{figure*}

\begin{figure*}
\includegraphics[width=17cm]{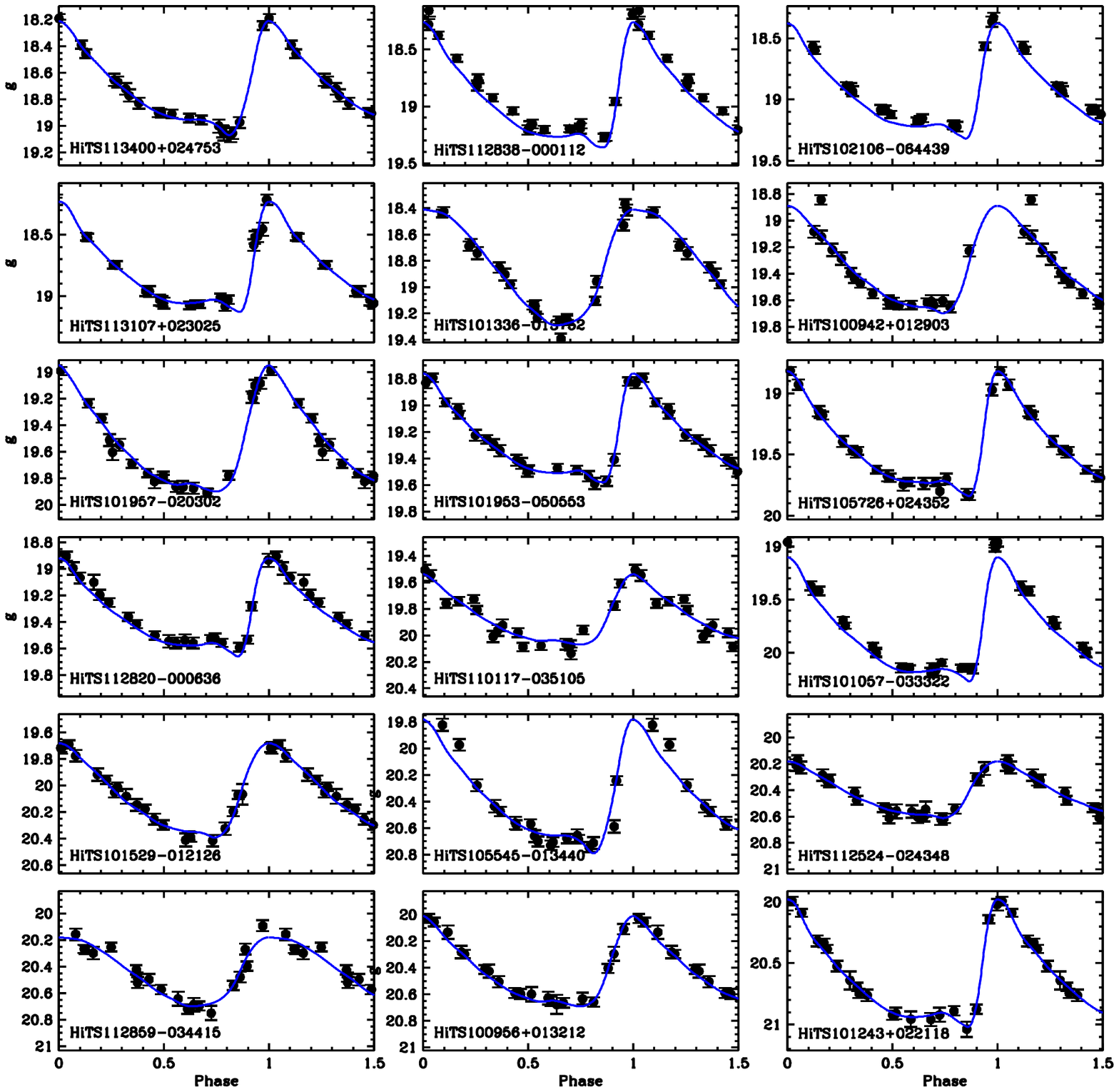}
\caption{Phased light curves of nearby ($<90$\,kpc) RR Lyrae stars in the field (4/4).}
\label{fig:lcField4}
\end{figure*}

Figures~\ref{fig:lcSextans1} to ~\ref{fig:lcSextans3} display the phased light curves of the RR Lyrae stars in the Sextans dwarf spheroidal galaxy. On the other hand, Figures~\ref{fig:lcField1}-\ref{fig:lcField4} show the light curves of the field RR Lyrae stars closer than $90$\,kpc.

\clearpage


\begin{thebibliography}{0}%
\makeatletter
\providecommand \@ifxundefined [1]{%
 \@ifx{#1\undefined}
}%
\providecommand \@ifnum [1]{%
 \ifnum #1\expandafter \@firstoftwo
 \else \expandafter \@secondoftwo
 \fi
}%
\providecommand \@ifx [1]{%
 \ifx #1\expandafter \@firstoftwo
 \else \expandafter \@secondoftwo
 \fi
}%
\providecommand \natexlab [1]{#1}%
\providecommand \enquote  [1]{``#1''}%
\providecommand \bibnamefont  [1]{#1}%
\providecommand \bibfnamefont [1]{#1}%
\providecommand \citenamefont [1]{#1}%
\providecommand \href@noop [0]{\@secondoftwo}%
\providecommand \href [0]{\begingroup \@sanitize@url \@href}%
\providecommand \@href[1]{\@@startlink{#1}\@@href}%
\providecommand \@@href[1]{\endgroup#1\@@endlink}%
\providecommand \@sanitize@url [0]{\catcode `\\12\catcode `\$12\catcode
  `\&12\catcode `\#12\catcode `\^12\catcode `\_12\catcode `\%12\relax}%
\providecommand \@@startlink[1]{}%
\providecommand \@@endlink[0]{}%
\providecommand \url  [0]{\begingroup\@sanitize@url \@url }%
\providecommand \@url [1]{\endgroup\@href {#1}{\urlprefix }}%
\providecommand \urlprefix  [0]{URL }%
\providecommand \Eprint [0]{\href }%
\providecommand \doibase [0]{http://dx.doi.org/}%
\providecommand \selectlanguage [0]{\@gobble}%
\providecommand \bibinfo  [0]{\@secondoftwo}%
\providecommand \bibfield  [0]{\@secondoftwo}%
\providecommand \translation [1]{[#1]}%
\providecommand \BibitemOpen [0]{}%
\providecommand \bibitemStop [0]{}%
\providecommand \bibitemNoStop [0]{.\EOS\space}%
\providecommand \EOS [0]{\spacefactor3000\relax}%
\providecommand \BibitemShut  [1]{\csname bibitem#1\endcsname}%
\let\auto@bib@innerbib\@empty
\end{thebibliography}%


\newpage

\begin{thebibliography}{}
\bibitem[Abadi et al.(2006)]{Abadi06} Abadi, M.~G., Navarro, J.~F., \& Steinmetz, M.\ 2006, \mnras, 365, 747 

\bibitem[Abbas et al.(2014)]{Abb14} Abbas, M.~A., Grebel, E.~K., Martin, N.~F., et al.\ 2014, \mnras, 441, 1230 

\bibitem[Abbott et al.(2015)]{abbott15} Abbott, T., Abdalla, F.~B., Alleksic, J. et al.\ 2016, \mnras, 460, 1270

\bibitem[Ablimit \& Zhao(2017)]{Ablimit2017} Ablimit, I., \& Zhao, G.\ 2017, \apj, 846, 10 

\bibitem[Akhter et al.(2012)]{Akh12} Akhter, S., Da Costa, G.~S., Keller, S.~C., et al.\ 2012, \apj, 756, 23 

\bibitem[Amigo et al.(2012)]{Amigo2012} Amigo Fuentes, P., Catelan, M., Zoccali, M., \& Pontificia Universidad Cat\'olica de Chile. Facultad de F\'isica. The Milky Way in the making : . : Clues into the formation history of our galaxy from time-resolved photometry of globular clusters and dwarf satellite galaxies. Santiago, Chile, 2012.

\bibitem[Baker \& Willman(2015)]{BakerWillman15} Baker, M., \& Willman, B.\ 2015, \aj, 150, 160 

\bibitem[Bell et al.(2008)]{Bell2008} Bell, E.~F., Zucker, D.~B., Belokurov, V., et al.\ 2008, \apj, 680, 295-311 

\bibitem[Belokurov et al.(2007)]{Belokurov07} Belokurov, V., Evans, N.~W., Irwin, M.~J., et al.\ 2007, \apj, 658, 337 

\bibitem[Bertin \& Arnouts(1996)]{Bert96} Bertin, E., \& Arnouts, S.\ 1996, \aaps, 117, 393 

\bibitem[Bochanski et al.(2014)]{Boch14} Bochanski, J.~J., Willman, B., Caldwell, N., et al.\ 2014, \apjl, 790, LL5 

\bibitem[Boettcher et al.(2013)]{Boettcher13} Boettcher, E., Willman, B., Fadely, R., et al.\ 2013, \aj, 146, 94 

\bibitem[Bonaca et al.(2012)]{Bonaca12} Bonaca, A., Juri{\'c}, M., Ivezi{\'c}, {\v Z}., et al.\ 2012, \aj, 143, 105 

\bibitem[Boubert et al.(2017)]{Boubert17} Boubert, D., Belokurov, V., Erkal, D., et al.\ 2017, arXiv:1711.07493 

\bibitem[Brown et al.(2005)]{Brown05} Brown, W.~R., Geller, M.~J., Kenyon, S.~J., et al.\ 2005, \apjl, 622, L33 

\bibitem[Bullock \& Johnston(2005)]{BJ05} Bullock, J.~S., \& Johnston, K.~V.\ 2005, \apj, 635, 931 

\bibitem[C{\'a}ceres \& Catelan(2008)]{Caceres08} C{\'a}ceres, C., \& Catelan, M.\ 2008, \apjs, 179, 242

\bibitem[Cacciari \& Clementini(2003)]{Cacc03} Cacciari, C., \& Clementini, G.\ 2003, in Stellar Candles for the Extragalactic Distance Scale, ed. D. Alloin \& W. Gieren (Lecture Notes in Physics, Vol. 635; Berlin: Springer), 105

\bibitem[Carlin et al.(2012)]{Carlin12} Carlin, J.~L., Yam, W., Casetti-Dinescu, D.~I., et al.\ 2012, \apj, 753, 145 

\bibitem[Carollo et al.(2007)]{Carollo07} Carollo, D., Beers, T.~C., Lee, Y.~S., et al.\ 2007, \nat, 450, 1020

\bibitem[Catelan et al.(2004)]{Catelan04} Catelan, M., Pritzl, B.~J., \& Smith, H.~A.\ 2004, \apjs, 154, 633 

\bibitem[Catelan 
\& Cort{\'e}s(2008)]{Cat08} Catelan, M., \& Cort{\'e}s, C.\ 2008, \apjl, 676, L135

\bibitem[Catelan(2009)]{Catelan2009} Catelan, M.\ 2009, \apss, 320, 261 

\bibitem[Catelan \& Smith(2015)]{Catelan2015} Catelan, M., \& Smith, H. ~A. \ 2015, Pulsating Stars (New York: Wiley, VCH)

\bibitem[Clementini(2010)]{Clem10} Clementini, G.\ 2010, in Variable Stars, the Galactic Halo and Galaxy Formation, ed. C. Sterken, N. Samus, \& L. Szabados (Russia: Sternberg Astronomical Institute of Moscow Univ.), 107

\bibitem[Clementini(2014)]{Clementini2014} Clementini, G. 2014, in Proc. IAU Symp. 301, Precision Asteroseismology,ed. J. A. Guzik et al. (Cambridge: Cambridge Univ. Press), 129

\bibitem[Cohen et al.(2015)]{Cohen15} Cohen, J.~G., Sesar, B., Banholzer, S., et al.\ 2015, arXiv:1509.05997 

\bibitem[Collins et al.(2016)]{Collins2016} Collins, M.~L.~M., Tollerud, E.~J., Sand, D.~J., et al.\ 2016, arXiv:1608.05710

\bibitem[Cooper et al.(2010)]{Cooper2010} Cooper, A.~P., Cole, S., Frenk, C.~S., et al.\ 2010, \mnras, 406, 744 

\bibitem[Cooper et al.(2013)]{Cooper2013} Cooper, A.~P., D'Souza, R., Kauffmann, G., et al.\ 2013, \mnras, 434, 3348 

\bibitem[Cort{\'e}s \& Catelan(2008)]{Cortes08} Cort{\'e}s, C., \& Catelan, M.\ 2008, \apjs, 177, 362-372 

\bibitem[Das et al.(2016)]{Das2016} Das, P., Williams, A., \& Binney, J.\ 2016, \mnras, 463, 3169 

\bibitem[De Propris et al.(2010)]{DePropris2010} De Propris, R., Harrison, C.~D., \& Mares, P.~J.\ 2010, \apj, 719, 1582 

\bibitem[Deason et al.(2011)]{Deas11} Deason, A.~J., Belokurov, V., \& Evans, N.~W.\ 2011, \mnras, 416, 2903 

\bibitem[Deason et al.(2013)]{Deas13} Deason, A.~J., Belokurov, V., Evans, N.~W., et al.\ 2013, \apj, 763, 113 

\bibitem[Deason et al.(2014)]{Deas14} Deason, A.~J., Belokurov, V., Koposov, S.~E., et al.\ 2014, \apj, 787, 30 

\bibitem[Demarque et al.(2000)]{Dem00} Demarque, P., Zinn, R., Lee, Y.-W., et al.\ 2000, \aj, 119, 1398 

\bibitem[Drake et al.(2013a)]{Dra13} Drake, A.~J., Catelan, M., Djorgovski, S.~G., et al.\ 2013, \apj, 763, 32 

\bibitem[Drake et al.(2013b)]{Dra13_100kpc} Drake, A.~J., Catelan, M., Djorgovski, S.~G., et al.\ 2013, \apj, 765, 154 

\bibitem[Drake et al.(2014)]{Dra14} Drake, A.~J., Graham, M.~J., Djorgovski, S.~G., et al.\ 2014, \apjs, 213, 9 

\bibitem[Drake et al.(2017)]{Drake17} Drake, A.~J., Djorgovski, S.~G., Catelan, M., et al.\ 2017, \mnras, 469, 3688

\bibitem[Duffau et al.(2006)]{Duffau06a} Duffau, S., Zinn, R., Vivas, A.~K., et al.\ 2006, \apjl, 636, L97 

\bibitem[Duffau et al.(2014)]{Duffau14} Duffau, S., Vivas, A.~K., Zinn, R., et al. 2014, \aap, 566, A118 

\bibitem[Eadie \& Harris(2016)]{Eadie2016} Eadie, G.~M., \& Harris, W.~E.\ 2016, \apj, 829, 108 

\bibitem[Fern{\'a}ndez-Trincado et al.(2015)]{Fernandez15} Fern{\'a}ndez-Trincado, J.~G., Vivas, A.~K., Mateu, C.~E., et al.\ 2015, \aap, 574, A15

\bibitem[Fiorentino et al.(2015)]{Fiorentino2015} Fiorentino, G., Bono, G., Monelli, M., et al.\ 2015, \apjl, 798, L12

\bibitem[Flaugher et al.(2015)]{Flaugher15} Flaugher, B., Diehl, H.~T., Honscheid, K., et al.\ 2015, \aj, 150, 150

\bibitem[Foreman-Mackey et al.(2013)]{Foreman2013} Foreman-Mackey, D., Hogg, D.~W., Lang, D., et al.\ 2013, \pasp, 125, 306 
 
\bibitem[F{\"o}rster et al.(2016)]{Forster16} F{\"o}rster, F., Maureira, J.~C., San Mart{\'{\i}}n, J., et al.\ 2016, \apj, 832, 155 

\bibitem[Garling et al.(2017)]{Garling17} Garling, C., Willman, B., Sand, D.~J., et al.\ 2017, arXiv:1711.07469

\bibitem[Gawiser et al.(2006)]{Gawiser2006} Gawiser, E., van Dokkum, P.~G., Herrera, D., et al.\ 2006, \apjs, 162, 1 

\bibitem[Geha et al.(2017)]{Geha2017} Geha, M., Wechsler, R.~H., Mao, Y.-Y., et al.\ 2017, \apj, 847, 4 

\bibitem[Grillmair(2006)]{Grillmair06_Orphan} Grillmair, C.~J.\ 2006, \apjl, 645, L37 

\bibitem[Grillmair \& Carlin(2016)]{GrillmairCarlin16} Grillmair, C.~J., \& Carlin, J.~L.\ 2016, Tidal Streams in the Local Group and Beyond, 420, 87 

\bibitem[Grillmair et al.(2015)]{Grillmair15_Orphan} Grillmair, C.~J., Hetherington, L., Carlberg, R.~G., et al.\ 2015, \apjl, 812, L26 

\bibitem[Hambly et al.(2001)]{Hambly2001} Hambly, N.~C., MacGillivray, H.~T., Read, M.~A., et al.\ 2001, \mnras, 326, 1279 

\bibitem[Hendel et al.(2017)]{Hendel17} Hendel, D., Scowcroft, V., Johnston, K.~V., et al.\ 2017, arXiv:1711.04663 

\bibitem[Ibata et al.(2001)]{Iba01} Ibata, R., Irwin, M., Lewis, G.~F., et al.\ 2001, \apjl, 547, L133 

\bibitem[Iorio et al.(2017)]{Iorio2017} Iorio, G., Belokurov, V., Erkal, D., et al.\ 2017, arXiv:1707.03833 

\bibitem[Irwin et al.(1990)]{Irwin1990} Irwin, M.~J., Bunclark, P.~S., Bridgeland, M.~T., et al.\ 1990, \mnras, 244, 16P 

\bibitem[Irwin \& Hatzidimitriou(1995)]{Irwin95} Irwin, M., \& Hatzidimitriou, D.\ 1995, \mnras, 277, 1354 

\bibitem[Ivezi{\'c} et al.(2005)]{Ive05} Ivezi{\'c}, {\v Z}., Vivas, A.~K., Lupton, R.~H., et al.\ 2005, \aj, 129, 1096 

\bibitem[Ivezi{\'c} et al.(2008)]{Ivezic08} Ivezi{\'c}, {\v Z}., Tyson, J.~A., Abel, B., et al.\ 2008, arXiv:0805.2366 

\bibitem[Johnston et al.(2012)]{Johnston12} Johnston, K.~V., Sheffield, A.~A., Majewski, S.~R., et al.\ 2012, \apj, 760, 95 

\bibitem[Juri{\'c} et al.(2008)]{Juric08} Juri{\'c}, M., Ivezi{\'c}, {\v Z}., Brooks, A., et al.\ 2008, \apj, 673, 864-914 


\bibitem[Keller et al.(2009)]{Keller09} Keller, S.~C., da Costa, G.~S., \& Prior, S.~L.\ 2009, \mnras, 394, 1045 


\bibitem[Keller et al.(2008)]{Keller08} Keller, S.~C., Murphy, S., Prior, S., et al.\ 2008, \apj, 678, 851-864 


\bibitem[Kirby et al.(2011)]{Kirby2011} Kirby, E.~N., Lanfranchi, G.~A., Simon, J.~D., et al.\ 2011, \apj, 727, 78 

\bibitem[Law \& Majewski(2010)]{Law10} Law, D.~R., \& Majewski, S.~R.\ 2010, \apj, 718, 1128 

\bibitem[Law \& Majewski(2016)]{LawMajewski16} Law, D.~R., \& Majewski, S.~R.\ 2016, Tidal Streams in the Local Group and Beyond, 420, 31 

\bibitem[Lee et al.(2003)]{Lee03} Lee, M.~G., Park, H.~S., Park, J.-H., et al.\ 2003, \aj, 126, 2840 

\bibitem[Lomb(1976)]{Lomb76} Lomb, N.~R.\ 1976, \apss, 39, 447 

\bibitem[Longmore et al.(1986)]{Longmore86} Longmore, A.~J., Fernley, J.~A., \& Jameson, R.~F.\ 1986, \mnras, 220, 279 

\bibitem[LSST Science Collaboration et al.(2009)]{LSST2009} LSST Science Collaboration, Abell, P.~A., Allison, J., et al.\ 2009, arXiv:0912.0201 

\bibitem[Majewski et al.(2003)]{Maj03} Majewski, S.~R., Skrutskie, M.~F., Weinberg, M.~D., et al.\ 2003, \apj, 599, 1082

\bibitem[Marconi(2009)]{Mar09} Marconi, M.\ 2009, American Institute of Physics Conference Series, 1170, 223 

\bibitem[Mart\'inez et al.(2018)]{Martinez2018} 
Mart\'inez, J., F{\"o}rster, F., Protopapas, P., et al.\ 2018, \apj, submitted

\bibitem[Mateo et al.(1991)]{Mateo91} Mateo, M., Nemec, J., Irwin, M., et al.\ 1991, \aj, 101, 892 

\bibitem[Mateo et al.(1995)]{Mateo1995} Mateo, M., Fischer, P., \& Krzeminski, W.\ 1995, \aj, 110, 2166 

\bibitem[Mateo(1998)]{Mateo98} Mateo, M.~L.\ 1998, \araa, 36, 435 

\bibitem[McConnachie(2012)]{McCon12} McConnachie, A.~W.\ 2012, \aj, 144, 4 

\bibitem[Medina et al.(2017)]{Medina2017} Medina, G.~E., Mu{\~n}oz, R.~R., Vivas, A.~K., et al.\ 2017, \apjl, 845, L10 

\bibitem[Moretti et al.(2009)]{Moretti2009} Moretti, M.~I., Dall'Ora, M., Ripepi, V., et al.\ 2009, \apjl, 699, L125 

\bibitem[Newberg et al.(2002)]{New02} Newberg, H.~J., Yanny, B., Rockosi, C., et al.\ 2002, \apj, 569, 245 

\bibitem[Newberg et al.(2007)]{Newberg07} Newberg, H.~J., Yanny, B., Cole, N., et al.\ 2007, \apj, 668, 221 

\bibitem[Newberg et al.(2010)]{Newberg10} Newberg, H.~J., Willett, B.~A., Yanny, B., et al.\ 2010, \apj, 711, 32 

\bibitem[Oluseyi et al.(2012)]{Oluseyi12} Oluseyi, H.~M., Becker, A.~C., Culliton, C., et al.\ 2012, \aj, 144, 9 

\bibitem[Oosterhoff(1939)]{Oosterhoff1939} Oosterhoff, P.~T.\ 1939, The Observatory, 62, 104

\bibitem[Palaversa et al.(2013)]{Pal13} Palaversa, L., 
Ivezi{\'c}, {\v Z}., Eyer, L., et al.\ 2013, \aj, 146, 101 

\bibitem[Pila-D{\'{\i}}ez et al.(2015)]{PilaDiez2015} Pila-D{\'{\i}}ez, B., de Jong, J.~T.~A., Kuijken, K., et al.\ 2015, \aap, 579, A38 

\bibitem[Pillepich et al.(2014)]{Pillepich14} Pillepich, A., Vogelsberger, M., Deason, A., et al.\ 2014, \mnras, 444, 237 

\bibitem[Prior et al.(2009)]{Prior09} Prior, S.~L., Da Costa, G.~S., \& Keller, S.~C.\ 2009, \apj, 704, 1327

\bibitem[Roderick et al.(2016)]{Roderick2016} Roderick, T.~A., Jerjen, H., Da Costa, G.~S., et al.\ 2016, \mnras,  460, 30  

\bibitem[Saha(1985)]{Saha85} Saha, A.\ 1985, \apj, 289, 310 

\bibitem[Sanderson(2016)]{Sanderson16} Sanderson, R.~E.\ 2016, \apj, 818, 41 

\bibitem[Sanderson et al.(2017)]{Sanderson2017b} Sanderson, R.~E., Secunda, A., Johnston, K.~V., et al.\ 2017, \mnras, 470, 5014 

\bibitem[Scargle(1982)]{Scar82} Scargle, J.~D.\ 1982, \apj, 263, 835 

\bibitem[Schlafly \& Finkbeiner(2011)]{Sch11} Schlafly, E.~F., \& Finkbeiner, D.~P.\ 2011, \apj, 737, 103 

\bibitem[Schlegel et al.(1998)]{SFD98} Schlegel, D.~J., Finkbeiner, D.~P., \& Davis, M.\ 1998, \apj, 500, 525 

\bibitem[Sesar et al.(2007)]{Ses07} Sesar, B., Ivezi{\'c}, {\v Z}., Lupton, R.~H., et al.\ 2007, \aj, 134, 2236 

\bibitem[Sesar et al.(2010)]{Ses10} Sesar, B., Ivezi{\'c}, {\v Z}., Grammer, S.~H., et al.\ 2010, \apj, 708, 717 

\bibitem[Sesar et al.(2011)]{Ses11} Sesar, B., Juri{\'c}, M., \& Ivezi{\'c}, {\v Z}.\ 2011, \apj, 731, 4 

\bibitem[Sesar et al.(2013)]{Sesar13a} Sesar, B., Ivezi{\'c}, {\v Z}., Stuart, J.~S., et al.\ 2013, \aj, 146, 21 

\bibitem[Sesar et al.(2013)]{Sesar13} Sesar, B., Grillmair, C.~J., Cohen, J.~G., et al.\ 2013, \apj, 776, 26 

\bibitem[Sesar et al.(2014)]{Ses14} Sesar, B., Banholzer, S.~R., Cohen, J.~G., et al.\ 2014, \apj, 793, 135 

\bibitem[Sesar et al.(2017a)]{Sesar2017a} Sesar, B., Hernitschek, N., Mitrovi{\'c}, S., et al.\ 2017a, \aj, 153, 204 

\bibitem[Sesar et al.(2017b)]{Sesar2017b} Sesar, B., Hernitschek, N., Dierickx, M.~I.~P., et al.\ 2017b, \apjl, 844, L4 

\bibitem[Simon \& Geha(2007)]{Simon2007} Simon, J.~D., \& Geha, M.\ 2007, \apj, 670, 313 

\bibitem[Slater et al.(2016)]{Slater2016} Slater, C.~T., Nidever, D.~L., Munn, J.~A., et al.\ 2016, \apj, 832, 206 

\bibitem[Tsujimoto et al.(1997)]{Tsu97} Tsujimoto, T., 
Miyamoto, M., \& Yoshii, Y.\ 1997, Hipparcos - Venice '97, 402, 639 

\bibitem[VanderPlas et al.(2012)]{Vander12} VanderPlas, J., Connolly, A.~J., Ivezic, Z., et al.\ 2012, in Proc. Conf. Intelligent Data Understanding (CIDU), ed. K. Das, N.V. Chawla, \& A. N. Srivastava (Boulder, CO: NCAR), 47 

\bibitem[VanderPlas \& Ivezi{\'c}(2015)]{VanderPlas15} VanderPlas, J.~T., \& Ivezi{\'c}, {\v Z}.\ 2015, \apj, 812, 18 

\bibitem[Valdes et al.(2014)]{Valdes2014} Valdes, F., Gruendl, R., \& DES Project 2014, Astronomical Data Analysis Software and Systems XXIII, 485, 379 

\bibitem[Vivas et al.(2001)]{Vivas01} Vivas, A.~K., Zinn, R., Andrews, P., et al.\ 2001, \apjl, 554, L33 

\bibitem[Vivas et al.(2004)]{Viv04} Vivas, A.~K., Zinn, R., Abad, C., et al.\ 2004, \aj, 127, 1158 

\bibitem[Vivas \& Zinn(2006)]{Vivas2006} Vivas, A.~K., \& Zinn, R.\ 2006, \aj, 132, 714 

\bibitem[Vivas et al.(2016)]{Viv16} Vivas, A.~K., Olsen, K., Blum, R., et al.\ 2016, \aj, 151, 118 

\bibitem[Vivas et al.(2016)]{Vivas2016b} Vivas, A.~K., Zinn, R., Farmer, J., et al.\ 2016, \apj, 831, 165

\bibitem[Vivas et al.(2017)]{Vivas17} Vivas, A.~K., Saha, A., Olsen, K., et al.\ 2017, \aj, 154, 85

\bibitem[Watkins et al.(2009)]{Wat09} Watkins, L.~L., Evans, N.~W., Belokurov, V., et al.\ 2009, \mnras, 398, 1757 

\bibitem[Watkins et al.(2010)]{Watkins10} Watkins, L.~L., Evans, N.~W., \& An, J.~H.\ 2010, \mnras, 406, 264 

\bibitem[Wetterer \& McGraw(1996)]{Wetterer1996} Wetterer, C.~J., \& McGraw, J.~T.\ 1996, \aj, 112, 1046 

\bibitem[Xue et al.(2015)]{Xue2015} Xue, X.-X., Rix, H.-W., Ma, Z., et al.\ 2015, \apj, 809, 144 

\bibitem[Zechmeister \& K\"{u}rster(2009)]{Zech09} Zechmeister, M., K\"{u}rster, M.\ 2009, \aap, 496, 577 

\bibitem[Zinn(1993)]{Zinn93} Zinn, R.\ 1993, The Globular Cluster-Galaxy Connection, ASP Conference Series, Vol 48, 38

\bibitem[Zinn et al.(2014)]{Zinn14} Zinn, R., Horowitz, B., Vivas, A.~K., et al.\ 2014, \apj, 781, 22 

\bibitem[Zolotov et al.(2009)]{Zolo09} Zolotov, A., Willman, B., Brooks, A.~M., et al.\ 2009, \apj, 702, 1058 

\end{thebibliography}
\end{document}